\documentclass[%
superscriptaddress,prab,10pt,
 amsmath,amssymb,
 aps,
twocolumn
]{revtex4-2}

\usepackage{graphicx}
\usepackage{dcolumn}
\usepackage{bm}
\usepackage{paralist}
\usepackage{booktabs}
\usepackage{caption}
\usepackage{epstopdf}
\usepackage{epsfig}
\usepackage{natbib}
\usepackage{textcase}
\usepackage{url}
\usepackage{makecell}
\usepackage{tabularx}
\usepackage{multirow}
\usepackage{array}
\usepackage{ragged2e}
\newcolumntype{P}[1]{>{\RaggedLeft\arraybackslash}p{#1}}
\newcolumntype{C}[1]{>{\centering\let\newline\\\arraybackslash\hspace{0pt}}m{#1}}
\usepackage{wrapfig}
\usepackage{float}
\usepackage[number-unit-separator=~]{siunitx} 
\usepackage{threeparttable}

\usepackage{textcomp}
\usepackage[]{hyperref}
\usepackage{xcolor}
\definecolor{APSBlue}{RGB}{46, 48, 146}
\hypersetup{
    colorlinks,
    linkcolor = APSBlue,
    citecolor = APSBlue,
    urlcolor = APSBlue
}
\usepackage{xurl}

\makeatletter\def\Hy@Warning#1{}\makeatother 
\usepackage{microtype} 
\usepackage{pifont}
\usepackage[english]{babel} 
\DeclareCaptionFormat{figures}{\justifying #1#2\quad{#3}}
\DeclareCaptionFormat{tables}{\justifying #1#2\quad{#3}}
\usepackage{placeins}
\begin{document}

\preprint{APS/123-QED} 

\title{Post-irradiation examination of a prototype tantalum-clad target \\
for the Beam Dump Facility at CERN}

\author{Tina Griesemer}
 \email{tina.griesemer@cern.ch}
\author{Rui Franqueira Ximenes}
 \email{rui.franqueira.ximenes@cern.ch}

\author{Gonzalo Arnau Izquierdo} 
\author{Ignacio Aviles Santillana} 
\affiliation{CERN, 1211 Geneva 23, Switzerland}

\author{Thomas Brehm}
\affiliation{Framatome, Erlangen, Germany}

\author{Adria Gallifa Terricabras}
\author{Stefan Höll}
\author{Richard Jacobsson}
\affiliation{CERN, 1211 Geneva 23, Switzerland}

\author{Marco Kaiser}
\author{Roman Kuchar}
\affiliation{Framatome, Erlangen, Germany}

\author{Ana Teresa Pérez Fontenla} 
\affiliation{CERN, 1211 Geneva 23, Switzerland}

\author{Alexey Rempel}
\affiliation{Framatome, Erlangen, Germany}

\author{Oscar Sacristan De Frutos} 
\affiliation{CERN, 1211 Geneva 23, Switzerland}

\author{Marcel Schienbein}
\affiliation{Framatome, Erlangen, Germany}

\author{Stefano Sgobba}
\author{Marco Calviani}
 \email{marco.calviani@cern.ch}
\affiliation{CERN, 1211 Geneva 23, Switzerland}

\date{\today}

\begin{abstract}
The Beam Dump Facility (BDF) is a planned new fixed-target installation in CERN's North Area that is expected to start operating by the end of the decade. A high-energy proton beam of \SI{400}{GeV/\textit{c}} will be delivered in \SI{1}{}-\unit{s} pulses of \SI{4e13}{} protons every \SI{7.2}{s}, amounting to \SI{4e19}{}~protons on target (PoT) per year and a resulting average thermal power deposition of \SI{305}{kW}. The experiment requires high-Z and high-density target materials and involves challenging thermomechanical conditions; hence, a core made of refractory metals is proposed in conjunction with a water-cooling circuit. Since direct contact of water with the target materials---tungsten and a molybdenum-based alloy, TZM---would induce erosion and corrosion, the target blocks are clad with a tantalum alloy by means of hot isostatic pressing. To verify the reliability of the design and manufacturing process of the BDF target, a reduced-scale prototype target was manufactured and irradiated with an accumulated \SI{2.4e16}{PoT} in 2018.
This paper presents the results and conclusions of the post-irradiation examination (PIE) conducted on this prototype BDF target \SI{2.5}{years} after its operation.
The PIE consisted of nondestructive and destructive testing activities, including film imaging, microscopy, metrology, ultrasonic testing, microstructural analysis, and mechanical and thermal characterization.
No effects of irradiation resulting from thermally induced stresses, such as geometrical changes or cracks in the cladding material, were observed, and it was determined that the diffusion bonding between the core and cladding materials was robust. Moreover, no changes were detected in the microstructures of the bulk materials or in the mechanical properties of the core materials.
In conclusion, the robustness of the target baseline design and manufacturing process was confirmed, validating its suitability for operating under the desired beam conditions. Nonetheless, observations regarding the brittleness of sintered tungsten indicate a potential area for improvement to enhance the lifetime of the BDF target. Overall, the examinations provided valuable insights into the performance of the prototype BDF target and indicated potential refinements for the future BDF target complex.
\end{abstract}

\maketitle

\let\clearpage\relax
\section{Introduction}
In research facilities such as neutron spallation sources, high-power particle-producing targets are often made of refractory metals due to their high density, high atomic and mass numbers, and outstanding thermomechanical properties. These materials have to withstand challenging thermomechanical conditions with high pulse intensities and high average power. In various facilities, such as KENS~\cite{kawai2001fabrication}, LANSCE~\cite{nelson2012fabrication}, ISIS~\cite{dey2018strategies}, and CSNS~\cite{wei2021advance}, a dedicated water-cooling system and tantalum (Ta) cladding are employed.

The Search for Hidden Particles (SHiP) collaboration is proposing a new fixed-target facility at CERN---the Beam Dump Facility (BDF)---to search for ``hidden'' particles such as weakly interacting long-lived particles and different types of light dark matter~\cite{Albanese:2878604}. The Super Proton Synchrotron (SPS) will deliver a high-energy beam to the experiment with proton momentum up to \SI{400}{GeV/\textit{c}}, \SI{4e13}{} protons per pulse (ppp), beam energies of up to \SI{2.6}{MJ/pulse} slowly extracted with a 1-s-long spill, a repetition rate of \SI{7.2}{s}, and a projected cumulative \SI{4e19}{} protons on target (PoT) per year. It is expected that this will result in an average power deposition in the target of approximately \SI{305}{kW}, with the pulse-average power reaching \SI{2.2}{MW}~\cite{ahdida2022ship}. The experiment requires target materials that have a high density, high atomic and mass numbers, and a short nuclear interaction length; as such, TZM (a molybdenum alloy with titanium and zirconium) and pure tungsten (W) have been selected.

\begin{figure}[t]
\centering
\includegraphics[width=\columnwidth]{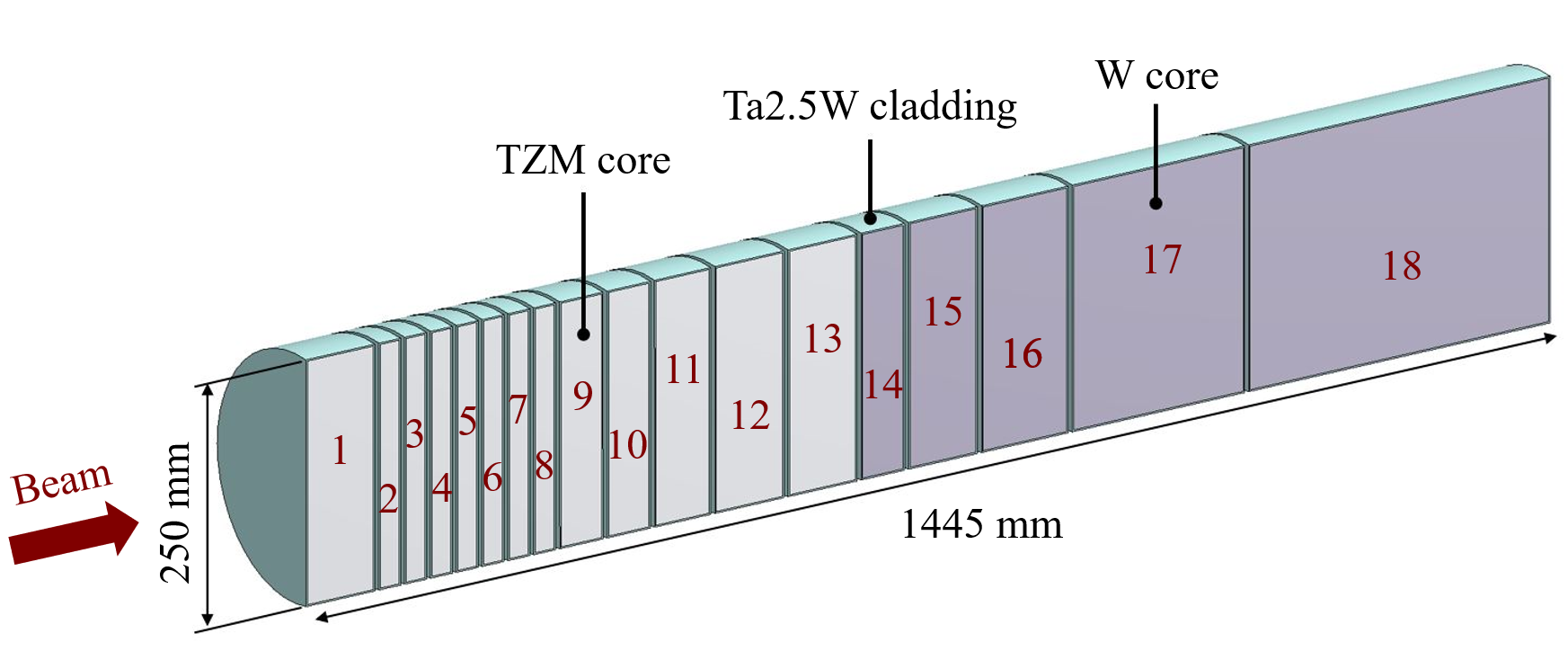}
\caption{Layout of the BDF target blocks including their numbering.}
\label{fig:target_blocks}
\end{figure}

Due to the high thermal power deposited by the beam, to remain within reasonable temperature limits, water cooling is needed for the target; however, refractory metals such as W experience erosion, corrosion, and embrittlement when in direct contact with water~\cite{lillard2000corrosion}. One way to prevent direct contact of the coolant with the target blocks while still allowing reliable heat transfer is to clad them with an erosion- and corrosion-resistant material. Tantalum was selected for this purpose due to its corrosion resistance and diffusion-bonding compatibility with both TZM and W~\cite{busom2020application}. The BDF target design consists of a stack of 18 target blocks with thicknesses in the range 25--\SI{350}{mm}, making a total length of \SI{1445}{mm}; each has a diameter of \SI{250}{mm}~\cite{ahdida2019sps} (see Fig.~\ref{fig:target_blocks}). There are 5-mm-wide water channels positioned between each pair of target blocks, and the thicknesses of the blocks were individually optimized according to the energy deposited by the beam in each section. The energy deposited in the target will be greater when using a higher-density material such as W; consequently, the first 13 blocks required the lower-density TZM as their core material to manage the thermal load, while the cores of the remaining five blocks were made of pure W. The beam is deflected radially by \SI{50}{mm} to impact the target in a circular motion, performing four turns during a single 1-s-long pulse.

To validate the design of the BDF target, a prototype was designed, manufactured, and tested with high-intensity proton beam at CERN. This prototype target had a smaller diameter of \SI{80}{mm}, but the other design parameters were kept the same. Ta and Ta2.5W (a Ta alloy with 2.5\% W) were used as cladding, and 50-\unit{\micro \meter}-thick Ta foil was used at some interfaces to aid the diffusion-bonding process~\cite{busom2020application}. Simulations were performed to define the beam parameters and ensure the similarity of the temperature and stress distributions of the prototype target and the actual BDF target.

In 2018, the prototype BDF target was irradiated on a dedicated test bench in the North Area at CERN~\cite{sola2019design,sola2019beam,franqueira2021jacow}. After a cool-down phase of \SI{1.5}{years}, it was opened to reveal the 18 reduced-scale target blocks. Based on the results of temperature and stress simulations, the six most critical blocks were selected for further analysis.

Since the BDF target will be exposed to a beam power of up to \SI{356}{kW}, a robust target design is crucial. Therefore, post-irradiation examination (PIE) investigations were conducted after the successful operation of the prototype BDF target. Previous PIE studies have included analysis of the microstructure and mechanical properties of W~\cite{horvath2018annealing,papadakis2021competing}, Ta alloys~\cite{byun2008dose}, and Ta-clad W~\cite{saito2023mechanical}. In the present study, the PIE assessed the reliability of the target design and manufacturing process while being operated under the intended beam conditions. In addition, the effects of new design features---such as using TZM as the target material, having Ta2.5W cladding, and using Ta foil at the bonding interfaces---were studied.

By means of both nondestructive and destructive tests, the PIE assessed the following aspects of the prototype target: (i)~the occurrence or absence of beam-induced thermomechanical effects; (ii)~the condition of the cladding surface, especially the presence or absence of cracks to assess whether the core is protected from water contact; (iii)~the comparative performance of the cladding materials Ta and Ta2.5W; (iv)~the reliability of the bonding between the core and cladding material; (v)~the advantages and disadvantages of using Ta foil to assist bonding; (vi)~any shortcomings of the design or manufacturing processes and any improvements that may be required.

\FloatBarrier
\section{Prototype target}
\subsection{Operation}
As the BDF experiment is expected to operate with the SPS proton beam, the North Area target zone at CERN (TCC2) was selected because it allows for slow extraction of the SPS beam. The TCC2 area lacks dilution magnets to generate the required beam pattern for the final facility; therefore, the beam parameters were tuned to match the thermomechanical conditions of the final BDF target (see Table~\ref{tab:beam_characteristics}). The resulting amounts of energy deposited on the targets were calculated using \textsc{fluka} Monte Carlo simulations~\cite{ahdida2022new} and imported into \textsc{ansys}\textsuperscript{\textregistered} {\footnotesize Mechanical\texttrademark}~\cite{ansys} for thermomechanical finite-element analysis (FEA). Intensities from \SI{3e12}{} to \SI{4e12}{ppp} were used to reproduce temperatures and stresses similar to those in the final target, with the highest intensity exceeding them (see Table~\ref{tab:target_temp_stress}); however, their distribution and stress evolution did not match due to the different beam parameters.

In 2018, the prototype BDF target was manufactured, assembled, and tested over three different days with approximately \SI{14}{h} of run time and an accumulated \SI{2.4e16}{PoT}; hence, the beam irradiation did not induce displacement damage on the target. These beam tests were instrumented with CERN's Beam TV (BTV) system, with one camera installed upstream (US) and another downstream (DS) of the target~\cite{Burger:1123677}. The instrumentation of the temperature and strain sensors on four blocks ran successfully during operation, and most of the sensors survived the testing conditions. In general, the measured data were coherent with the FEA results. During operation, one temperature sensor, on Block~8, exhibited higher values than predicted by FEA, and this therefore required further investigation in the PIE. In addition, a visual inspection during the intervention revealed surface discoloration in the beam-impact area~\cite{sola2019beam}.

\begin{table}[htbp]
   \centering
   \footnotesize
   \renewcommand{\arraystretch}{1.2}
   \caption{Beam parameters for the final BDF target compared to those for the prototype BDF target~\cite{sola2019beam}.}

   \begin{tabular}{l*{4}{c}}
       \toprule
       \toprule
                                && Final && Prototype \\
        Baseline characteristics && BDF target && BDF target \\
        \midrule
        Proton momentum (GeV/\textit{c})                                    && 400 &&   400\\
        Beam intensity (p$^+$/cycle)                && \SI{4e13}{} &&   3--\SI{4e12}{}\\
        Beam dilution                                              && Yes &&   No\\
        Horiz./vert. beam spot size (mm)                                  && 8/8 &&   3/2.5\\
        Cycle length (s)                                           && 7.2 &&  7.2\\
        Spill duration (s)                                         && 1.0 &&  1.0\\
        Average beam power (kW)                                    && 356 &&  35\\
        Average power on target (kW)                               && 305 &&  23\\
        Average beam power during                                  && 2.56 &&   0.26\\
        \hspace{3mm} spill (MW) &&\\
        Power density per spill (MW/m$^3$)           && 38 &&  38\\
       \bottomrule
       \bottomrule
   \end{tabular}
   \label{tab:beam_characteristics}
\end{table}

\begin{table}[htbp]
   \centering
   \footnotesize
   \renewcommand{\arraystretch}{1.2}
   \caption{Maximum temperature and maximum equivalent von Mises stress in the final BDF target in comparison with the prototype BDF target for a range of intensities~\cite{sola2019beam}.}
   \begin{tabular}{l*{10}{c}}
       \toprule
       \toprule
                &&\multicolumn{3}{c}{Maximum expected} &&&\multicolumn{3}{c}{Maximum expected}\\
                &&\multicolumn{3}{c}{temperature (\SI{}{\degree C})} &&&\multicolumn{3}{c}{stress (MPa)}\\\cline{3-5} \cline{8-10}
                && Final && Prototype target &&& Final && Prototype target \\
        Material  && target && 3--\SI{4e12}{ppp} &&& target && 3--\SI{4e12}{ppp}\\
        \midrule
        TZM    && 180 &&   240--300 &&& 130 &&   145--195\\
        W      && 150 &&   135--165 &&& 95 &&   85--110\\
        Ta2.5W && 160 &&   230--285 &&& 95 &&   85--120\\
       \bottomrule
       \bottomrule
   \end{tabular}
   \label{tab:target_temp_stress}
\end{table}

\begin{table*}[htbp]
   \centering
   \renewcommand{\arraystretch}{1.2}
   \setlength{\tabcolsep}{6pt}
   \caption{Summary of the blocks examined, listing their material composition, dimensions, and irradiation levels.}
   \begin{tabular}{l*{8}{c}}
       \toprule
       \toprule
        Block ID  &   3   & 4   &  8 & 9 & 14 & 15 & A & B \\
        \midrule
        Core material & TZM & TZM & TZM & TZM & W & W & TZM & W \\
        Cladding material & Ta2.5W & Ta2.5W & Ta & Ta2.5W & Ta & Ta & Ta & Ta\\
        Ta interlayer & -- & Downstream & -- & -- & Downstream & -- & -- & -- \\
        Diameter (mm) & 80 & 80 & 80 & 80 & 80 & 80 & 80 & 80 \\
        Thickness (mm) & 25 & 25 & 25 & 50 & 50 & 80 & 25 & 50 \\
        Irradiated & Yes & Yes & Yes & Yes & Yes & Yes & No & No\\
        Instrumented & No & Yes & Yes & Yes & Yes & No & No & No\\
        Dose rate (mSv/h) & 0.967 & 1.20 & 1.83 & 2.38 & 3.18 & 2.37 & -- & -- \\
        in July 2021 & & & & & & & & \\
       \bottomrule
       \bottomrule
   \end{tabular}
   \label{tab:overview_blocks}
\end{table*}

\begin{table*}[htbp]
   \centering
   \renewcommand{\arraystretch}{1.2}
   \setlength{\tabcolsep}{6pt}
   \caption{Overview of the type, number, and objective of each extracted specimen.}
   \begin{tabular}{p{3.3cm}C{0.7cm}C{0.7cm}C{0.7cm}C	{0.7cm}p{9cm}}
       \toprule
       \toprule
        Specimen type & \multicolumn{4}{c}{Extracted/analyzed samples} & Characteristics and objective\\
         & A & 4 & B & 14 &   \\
        \midrule
        Microsection (M) & 1/1 & 1/1 & 2/1 & 2/2 & Determining microstructural changes before and after irradiation.\\
        Flat tensile (T) & 6/4 & 6/4 & -- & 6/4 & Extraction in the region of highest stresses of the core materials.\\
        Shear (S) & 2/2 & 2/2 & 2/2 & 2/1 & Assessing the shear strength of the cladding--core bonding interface.\\
        Thermal diffusivity (D) & 6/0 & 6/6 & -- & 6/6 & Measuring the thermal resistance of the cladding--core interface.\\
       \bottomrule
       \bottomrule
    \end{tabular}
   \label{tab:overview_extraction}
\end{table*}

\subsection{Dismantling}
In January 2020, a fully remote intervention was conducted to extract the highly activated blocks of the prototype BDF target~\cite{franqueira2021jacow}. This involved removing and opening the prototype target, extracting some of the target blocks, and storing the remainder of the radioactive device. First, the cooling circuit was flushed to reduce the risk of residual contamination. CERN's Teodor\textsuperscript{\textregistered} and CERNBot robots~\cite{DiCastro:ICALEPCS2017-TUPHA127}, along with an overhead crane, were used after extensive mock-up testing. After opening the shielding, the plug-in system was unscrewed, and the prototype was taken to a bunker. All cabling connectors were detached, and stagnant water was drained. The tank was opened by unscrewing the downstream flange, the instrumentation wires connected to the flange were cut, and the half-shells containing the irradiated target blocks were pulled out. The upper half-shell was removed, and radiation-dose and contamination measurements were performed, revealing dose rates up to \SI{90}{mSv/h} after \SI{1.5}{years} of cool-down. The target blocks were marked with a pen to indicate their angular orientation with respect to the beam. Six target blocks were extracted with vacuum clamps and stored in a shielding container for the subsequent PIE studies.

\section{Sample extraction and testing}
This section describes the nondestructive and destructive testing methods that were used to evaluate the irradiated prototype BDF target; these methods were defined once the first observations had been conducted after the removal of the prototype from the beam line~\cite{franqueira2021jacow}. The PIE sought to evaluate the reliability of the current design and manufacturing process of the BDF target~\cite{ahdida2019sps}. Although the nondestructive tests were carried out on all six irradiated target blocks, the destructive tests were only performed on Block~4 (TZM) and Block~14 (W) because these were most critical in terms of the temperatures and stresses to which the core and cladding materials were subjected during beam operation~\cite{ahdida2019sps}. In addition, two unirradiated blocks from the same manufacturing batch as the prototype BDF target were used to demonstrate the methodology and testing parameters of the applied techniques. Table~\ref{tab:overview_blocks} lists all of the inspected blocks alongside their material compositions, dimensions, and irradiation levels.

\begin{figure}[htbp]
\centering
\includegraphics[width=.6\columnwidth]{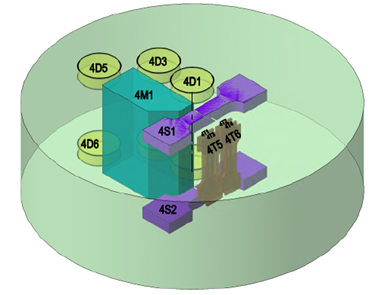}
\caption{Schematic of sample extraction from Block~4, showing one microsection (M), and six tensile (T), two shear (S), and six thermal-diffusivity (D) specimens.}
\label{fig:extraction_block}
\end{figure}

Specimens were extracted from the blocks by electric discharge machining (EDM) using a brass wire with a diameter of \SI{0.2}{mm}. Table~\ref{tab:overview_extraction} lists each of these specimens and their purposes. As an example, Fig.~\ref{fig:extraction_block} shows a schematic of the sampling of Block~4, which took into account both the beam-impact location determined through film imaging and the results from simulations~\cite{sola2019beam}. The tensile and shear specimens were taken from the regions of highest stress. It was found that the TZM presented the highest equivalent von Mises stresses \SI{8}{mm} from the beam impact, while W had the highest maximum principal stresses \SI{19}{mm} from the beam impact. At the bonding interfaces, the highest shear stresses were found \SI{4}{mm} from the beam impact in Block~4 and \SI{6.5}{mm} from the beam impact in Block~14. Thermal-diffusivity specimens were taken at different distances from the beam impact. Microsection specimens were extracted along the beam axis, covering all peak-stress regions in the radial direction. Specimens were first taken from the unirradiated blocks~A and B. After assessing the EDM cutting technique and the results of mechanical testing, blocks~4 and 14 were processed. Sample 14M2 (W) showed the highest dose rate of all the specimens, with \SI{480}{\micro Sv/h} at contact when measured in September 2023.

\subsection{Nondestructive tests}
\subsubsection{Film imaging}
The location of the center of the beam impact defines the areas of interest for subsequent testing; hence, film dosimetry was used to complement the data obtained from the BTV system. In this technique, specialized films exhibit a color change upon exposure to radiation; the higher the absorbed dose of ionizing radiation, the darker the color. On each flat surface, US and DS of each of the six irradiated blocks, a premarked Gafchromic\texttrademark\ film~\cite{gafcromic} was exposed for \SI{48}{h}. After this, each film was digitized and postprocessed based on the pixel intensities; however, the accuracy of this method was limited by the occurrence of rotational errors. The alignment of the films to the blocks was performed manually by a manipulator arm while relying on pen markings, and the subsequent positioning of the films in the scanner was also conducted manually. An additional visual verification step was applied to ensure that the circles detected during postprocessing matched each block's outer circle.

\subsubsection{Microscopy and metrology}
\begin{figure}[htbp]
\centering
\includegraphics[width=.8\columnwidth]{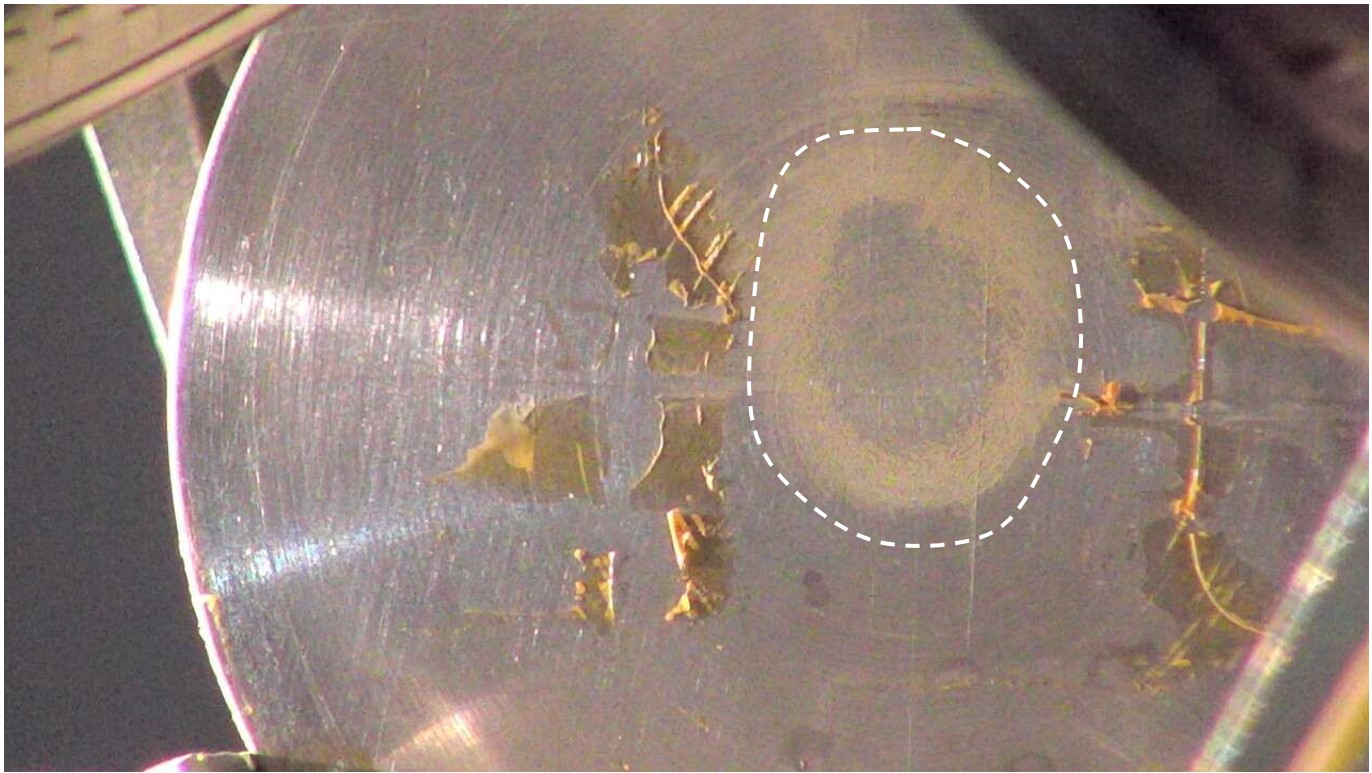}
\caption{Discoloration visible in the beam-impact area observed during the removal of the prototype target.}
\label{fig:discoloration}
\end{figure}

Microscopy and metrology were used to assess the presence of beam-induced damage on the target blocks, such as surface modifications and geometrical changes. In addition, the discoloration seen in the beam-impact region during the removal of the target blocks was analyzed (see Fig.~\ref{fig:discoloration}). High-resolution images were taken of all flat surfaces using a Keyence\textsuperscript{\textregistered} VHX-6000 digital microscope. The outer diameters were assessed using a caliper with a resolution of \SI{0.01}{mm}, obtaining 12 measurements from each block. For the flat surfaces, a contactless 3D optical macroscope---a Keyence VR-3100---was used to evaluate the global and local geometry. The global geometry was captured by panoramic images, and the line roughness was measured at three locations covering the center, mid-radius, and outer-radius positions. The results were compared against the parameters defined in the technical drawings: a diameter of $80 \pm 0.1$~\unit{mm} and an arithmetic average roughness (Ra) of \SI{1.6}{\micro \metre}.

Three approaches were used to determine the origin of the discoloration shown in Fig.~\ref{fig:discoloration}. First, deposit was extracted from the discolored center region with carbon tape (C-tape) attached to a pin-specimen holder. The chemical composition of this deposit was evaluated using an energy-dispersive X-ray spectroscopy (EDS) system integrated into a scanning electron microscopy (SEM) system. Second, SEM was conducted with EDS analysis to determine the chemical composition of apparent spots in the discolored regions using an acceleration voltage of \SI{20}{kV}. Third, because carbon deposition was suspected to be the cause of the discoloration, low-voltage EDS operating at \SI{5}{kV} was employed. The carbon content of Block~3 was determined in both the center region displaying discoloration and in the outer region without discoloration; the intention of the latter was to examine areas without specific features or particles. Subsequently, the results were compared.

\subsubsection{Ultrasonic testing}
Ultrasonic testing (UT) was performed to assess the bonding quality of the interfaces between the cladding and core materials. This examination included all flat surfaces (US and DS) of the two unirradiated and six irradiated blocks. An Olympus OmniScan MX2 was used with a linear pulser-receiver phased-array probe with inspection oil as a couplant. The equipment was calibrated according to EN~ISO~18563-3:2015, including two-point calibration using two dedicated reference blocks made of Ta. Three reference blocks made of Ta containing numerous flat bottom holes were used to define the detection threshold for insufficient bonding. The signal-to-noise ratio being above \SI{6}{dB} was taken to indicate delamination at the bonding interfaces. 

\subsection{Destructive tests}
\subsubsection{Microstructural examination}
The purpose of the microstructure specimens was to conduct examinations of the bulk materials and their interfaces, testing their hardness and analyzing them for the presence of any beam-induced damage. For preparation, samples AM1, BM1, 4M1, 14M1, and 14M2 were hot-mounted and polished. The polishing process comprised SiC grit, interim etching (60~s), diamond paste (30~s), suspension etching (15~s), and oxide polish suspension.

The regions of interest were along the beam axis, the US and DS interfaces, the core center, and the areas of highest core stresses and interface shear stresses. In these locations, light optical microscopy (LOM), SEM, EDS, and electron backscatter diffraction (EBSD) examinations were performed. The EBSD analysis included both initial broad scans and detailed mapping of a smaller number of locations. To determine the effects of irradiation, the EBSD mappings were evaluated by means of kernel average misorientation (KAM) and disorientation angle distributions.

To assess the mutual diffusion of cladding and core material after hot isostatic pressing (HIPing), the Ta/W interface of BM1 was investigated. EDS mapping was performed to determine the distributions of Ta and W in the interface region. In addition, a line scan perpendicular to the interface was conducted in the same region. The diameter of the electron beam is not precisely known; however, a diameter of less than \SI{10}{nm} is expected for the ZEISS Crossbeam 550 field-emission-gun SEM at \SI{5}{kV}. Using \textsc{casino}~\cite{casino2007}, Monte Carlo simulations were carried out to calculate the Ta signal distributions at a Ta/W interface based on electron-beam diameters of 1, 10, 20, 100, and \SI{200}{nm}.

Because it was suspected that the cracks discovered in 14M1 (W) would grow when applying destructive methods, only nondestructive methods were used for analysis of this sample. It was not feasible to apply UT after the extraction of the microsection specimen 14M1; as such, dye penetrant testing (PT) was applied to the polished side of the mounted 14M1 sample to confirm the detectability of cracks. After this, BM1 (W) (which was removed from the mount) and BM2 (W) were also tested with PT. In each case, the penetrant was applied for 30~min, the sample was rinsed with water, and the developer had a dwell time of 60~min.

\subsubsection{Indentation hardness testing}
Vickers hardness testing (HV0.5) was conducted with a dwell time of 10~s. Three profiles along the axial orientation were chosen: the beam axis; along the maximum shear stress in the interface; and along the maximum core stresses for each block. The cladding and core materials were subjected to five and ten equidistant indents, respectively. In the event of a recrystallized microstructure, an evaluation of the grain size in the core material was performed on the basis of the ASTM~E112 standard.

\subsubsection{Mechanical testing}
Mechanical tests were conducted to characterize the bulk core materials, the shear stresses at the cladding--core interfaces, and beam-induced thermomechanical effects on the mechanical properties. Tensile testing was performed on four specimens---two at 22\,\unit{\degreeCelsius} and two at 200\,\unit{\degreeCelsius}---extracted from each of blocks~A, 4, and 14, and the results were compared with previous test results, as listed in Table~\ref{tab:FH_mechanical_results}. All the shear specimens from blocks~A, B, 4, and 14 were tested at 22\,\unit{\degreeCelsius}. To confirm the testing parameters, unirradiated tensile specimens and shear specimens were assessed first, followed by the irradiated specimens. The EN~ISO~6892 standard was employed while using a strain rate of \SI{2.5e-4}{s^{-1}}.

To assess the breaking strength of the shear specimens in case of shear fractures, the ultimate shear strength (USS) of each bulk material was derived. A simplified approach to obtaining the USS is to multiply of the ultimate tensile strength (UTS) by \SI{0.577}{}~\cite{budynasshigley}, which resulted in USS values at 22\,\unit{\degreeCelsius} of 150, 215, 356, and \SI{104}{MPa} for Ta, Ta2.5W, TZM, and W, respectively~\cite{FHinternal2017}.

\begin{table}[htbp]
   \centering
   \renewcommand{\arraystretch}{1.2}
   \setlength{\tabcolsep}{2pt}
   \caption{Results of mechanical testing of the BDF bulk materials~\cite{FHinternal2017}, showing the yield strength (YS) and ultimate tensile strength (UTS) in MPa.}

    \begin{threeparttable}
    \begin{tabular}{l*{11}{c}}
       \toprule
       \toprule
       & \multicolumn{2}{c}{Ta\tnote{a}} &&\multicolumn{2}{c}{Ta2.5W\tnote{a}} && \multicolumn{2}{c}{TZM\tnote{b}} && \multicolumn{2}{c}{W\tnote{b}} \\ \cline{2-3}  \cline{5-6} \cline{8-9} \cline{11-12}
            & YS & UTS &&  YS & UTS &&  YS & UTS &&  YS & UTS \\
        \midrule
           22\,\unit{\degreeCelsius}       & \SI{210}{} & \SI{257}{}  && \SI{278}{} & \SI{371}{} && \SI{614}{} & \SI{617}{}    && -- & \SI{181}{}  \\
          200\,\unit{\degreeCelsius} & \SI{121}{} &\SI{231}{}   && \SI{227}{} &\SI{331}{}  && \SI{460}{} &\SI{551}{}     && -- &\SI{142}{} \\
       \bottomrule
       \bottomrule
    \end{tabular}
    \vspace{1mm}
    \begin{tablenotes}
        \footnotesize
        \item [a]Annealed sheets.
        \item [b]HIPed blocks taken from same manufacturing batch as BDF prototype target blocks.
    \end{tablenotes}
    \end{threeparttable}
    \label{tab:FH_mechanical_results}
\end{table}

\subsubsection{Thermal testing}
To determine the quality of the diffusion bonding, two-layered thermal-diffusivity specimens containing cladding and core materials were taken from Block~4 (Ta2.5W/TZM) and Block~14 (Ta/W). Six specimens were examined from each block; regarding their numbering, even (odd) sample numbers indicate the presence (absence) of a 50-\unit{\micro \meter}-thick Ta interlayer at the interface. The specimens each had a diameter of \SI{10}{mm} and a thickness of \SI{3}{mm}, and the interfaces were located approximately in their centers.

To determine the precise layer thicknesses of each sample, density measurements were performed using a Sartorius Quintix\textsuperscript{\textregistered}~224-1x scale with ethanol. The layer thicknesses were derived using the mixed density
and the total thickness.
A thin graphite coating was applied 
before laser flash analyses were carried out using a NETZSCH LFA~457 MicroFlash\textsuperscript{\textregistered} with a mercury-cadmium-telluride infrared detector, a laser voltage of 2210--2594~V, a helium atmosphere, and a temperature range of 30--450\,\unit{\degreeCelsius}. This allowed the thermal contact resistances ($R$-values) of each sample to be assessed; the thermal contact conductance (TCC) values, which are the reciprocal of the $R$-values, could then be examined.

A TCC sensitivity analysis was conducted by undertaking thermomechanical calculations using the FEA models for the Ta2.5W-clad blocks~4 and 14, as presented in the description of the specimen-extraction procedure. The TCC threshold for each block was established based on a 10\,\unit{\degreeCelsius} temperature rise compared to the ideal bonding conditions, which would correspond to an $R$-value of zero.

\FloatBarrier
\section{Results}
\subsection{Nondestructive tests}
\subsubsection{Film imaging}
\label{ch:res_film}
\begin{figure}[htbp]
\centering
\includegraphics[width=\columnwidth]{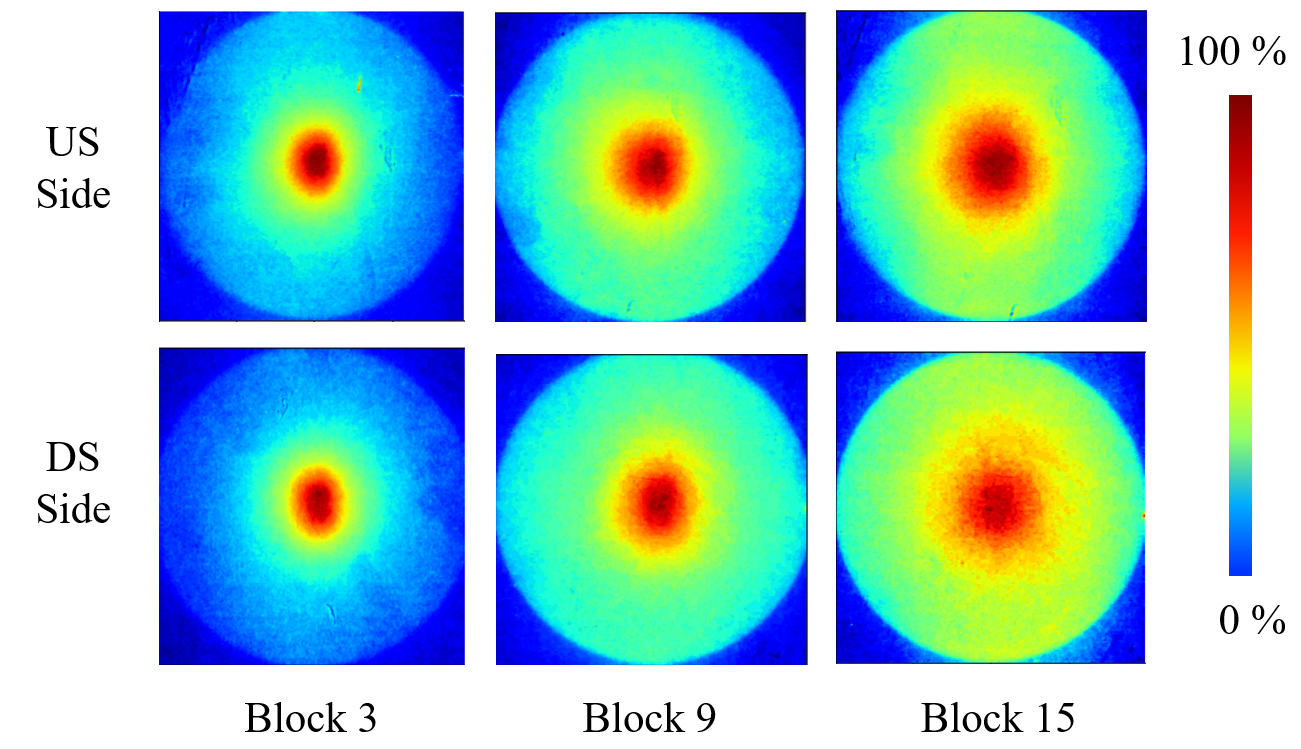}
\caption{Heat maps of Gafchromic films displaying the activation levels of each block. The scale of the legend is applied to each block individually. The DS-side films are displayed horizontally mirrored.}
\label{fig:gafcromic_postprocessed}
\end{figure}

\begin{figure}[htbp]
\centering
\includegraphics[width=\columnwidth]{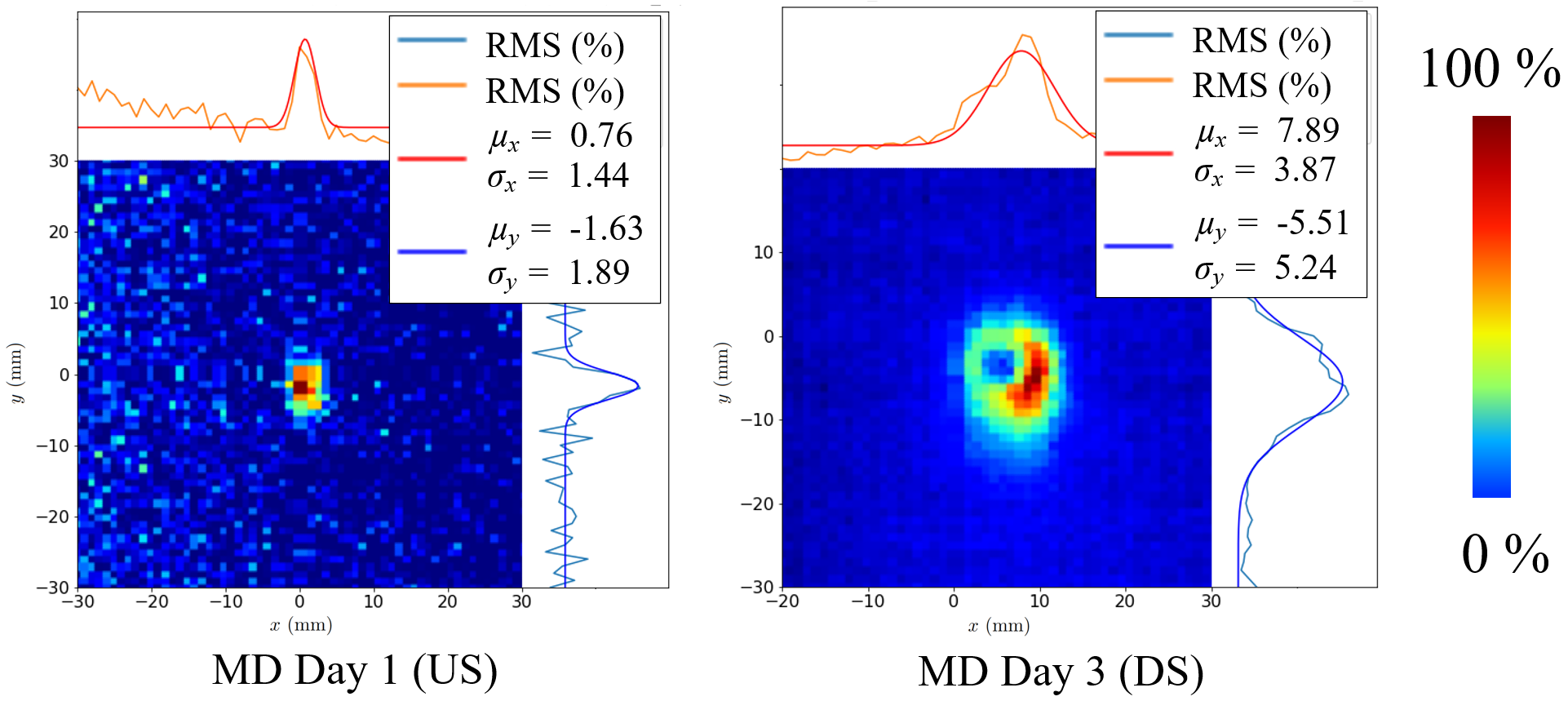}
\caption{Heat maps of the accumulated protons per BTV camera and day and their Gaussian fit parameters over both axes.}
\label{fig:BTV_postprocessed}
\end{figure}

\begin{figure}[htbp]
\centering
\includegraphics[width=\columnwidth]{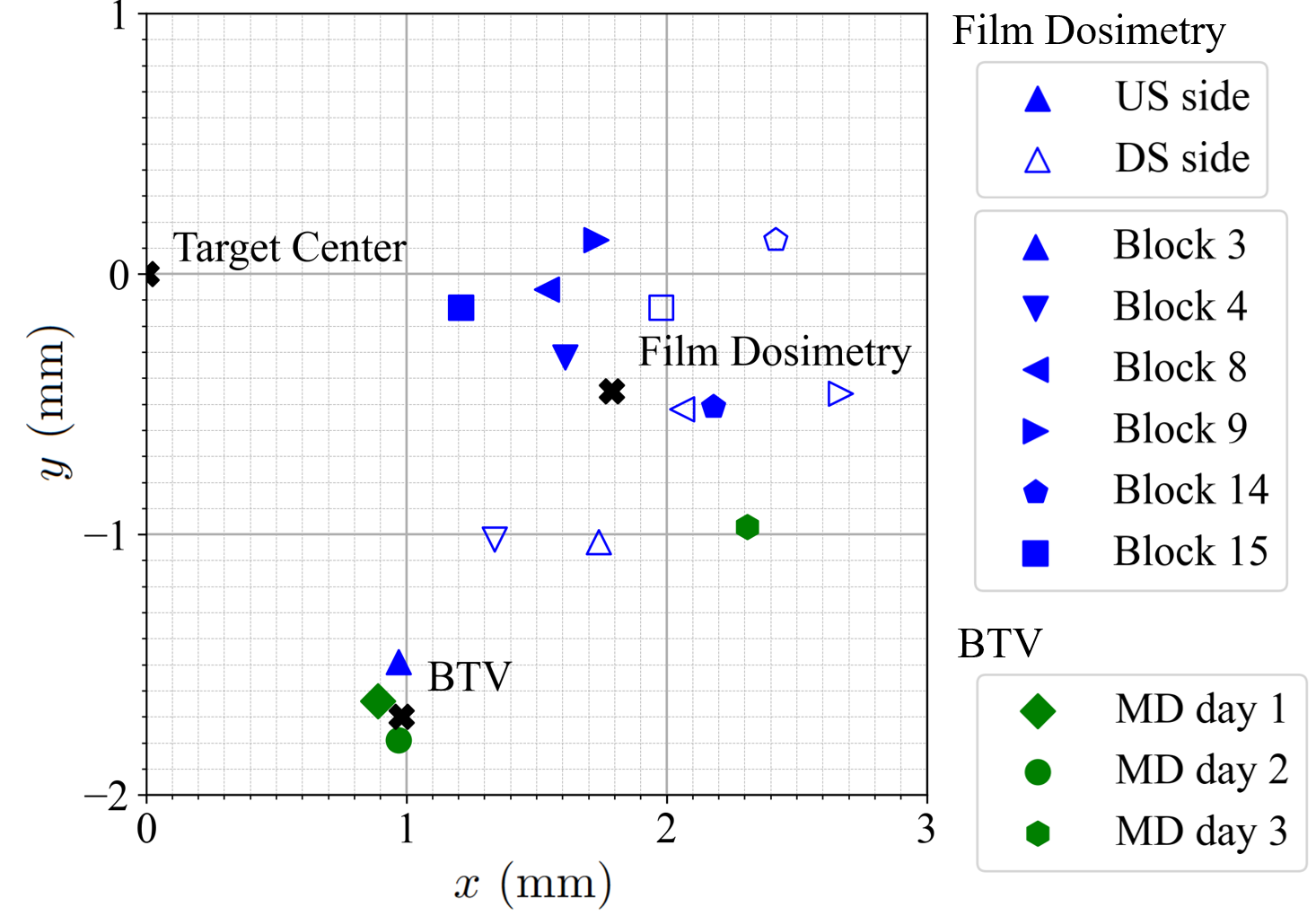}
\caption{Overview of the beam-impact locations from the film dosimetry and the BTV data with \textcolor{black}{\ding{54}} representing the determined average values and the center of the target block.}
\label{fig:beam offset}
\end{figure}

Figure~\ref{fig:gafcromic_postprocessed} shows the Gafchromic film results, and Fig.~\ref{fig:BTV_postprocessed} shows the data collected on two machine development (MD) days after postprocessing. The color scheme of each image is based on its respective maximum and minimum values. From the film imaging, beam diffusion from Block~3 to Block~15 can be observed, along with a slight but consistent beam offset from the block center toward the right-bottom side, resulting in an average beam offset of (1.8~mm, $-$0.5~mm). On all MD~days, the US camera displayed higher measurement accuracy than the DS camera, which exhibited a defocused beam spot without a Gaussian beam distribution. The beam profile monitored with the US camera was vertically elongated, and this matched the shapes observed on the Gafchromic films. The US camera failed on MD~day~2, and the DS camera failed on MD~day~1. The beam offset was determined solely by the US camera, and an approximation was applied for MD~day~2. The beam offset for MD~day~2 was derived based on the correlation between the DS camera and the US camera on MD~day~3. The BTV beam offset was determined by considering the share of PoT for each MD~day (38\%, 59\%, and 3\%, respectively), resulting in an offset of (1.0~mm, $-$1.7~mm). Figure~\ref{fig:beam offset} shows a plot including all of these beam offsets, with \textcolor{black}{\ding{54}} symbols indicating the target center and the determined average values. Both final beam offsets were located downward and to the right of the target center, and they were separated by approximately \SI{1}{mm} in their horizontal and vertical coordinates.

\subsubsection{Microscopy and metrology}
All caliper, topography, and roughness measurements were found to be within the manufacturing tolerances, and no geometrical changes were detected between the irradiated and unirradiated blocks. Topography measurements were performed on the flat surfaces, and concave morphologies were revealed in all the central areas, except for the DS~side of Block~4, which displayed a convexity of less than \SI{5}{\micro\meter} caused by machining. Therefore, swelling of the blocks was excluded. The roughness increased from the outer edge of the surface toward the center, and the expected Ra value of \SI{1.6}{\micro\meter} was mostly exceeded in the center region but was below this value in other areas.

\begin{figure}[htbp]
\centering
\includegraphics[width=\columnwidth]{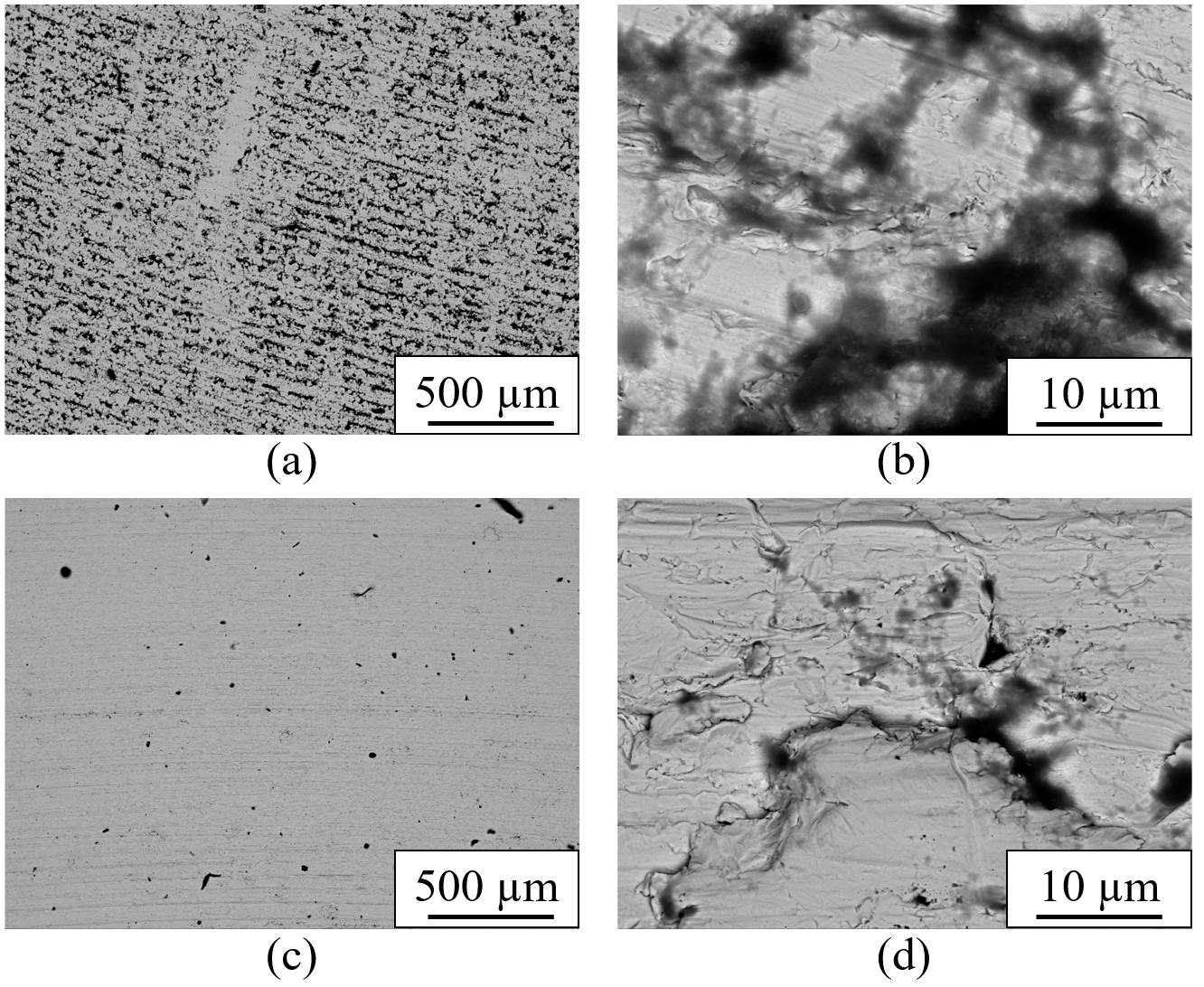}
\caption{Light optical microscopy images taken in the (a),(b)~central region and (c),(d)~outer area on the DS~side of Block~3.}
\label{fig:SEM_deposit}
\end{figure}

The LOM images of all flat surfaces confirmed the absence of any defects such as cracks, melting, separations, or pores caused by the beam impact; however, the turning grooves in the central region were more strongly pronounced, and lubricant residue was detected. Examination of the discoloration in the center areas revealed that the previously observed orange-like color was caused only by the lighting used during dismantling, and dark gray was seen in the LOM images. Block~3 presented the strongest discoloration, and it was thus analyzed further. The deposit that was extracted from the discolored region using C-tape exhibited multiple elements, with no particular element being clearly prominent. The SEM images of the DS face of Block~3 (Fig.~\ref{fig:SEM_deposit}) show greater concentrations of deposits in the turning grooves, matching the surface topography. At a higher magnification, spots with a grayish-to-blackish color arranged in an irregular network were observed, and they were denser and more pronounced in the central region. Especially at surface irregularities such as separations, material overlaps, or scratches, dense deposits were detected in all regions. EDS analyses were performed, focusing on apparent spots that showed carbon and nitrogen as the main elements. Areas in the central and outer regions that exhibited no evident spots were also investigated. Using a low acceleration voltage of \SI{5}{kV}, it was found that the area of discoloration in the center had a carbon concentration of approximately \SI{1.13}{wt\,\%}, and the outer region had a carbon concentration of approximately \SI{1.32}{wt\,\%}. In addition to carbon, oxygen and tantalum were detected, but nitrogen was not present.

\begin{figure}[htbp]
\centering
\includegraphics[width=\columnwidth]{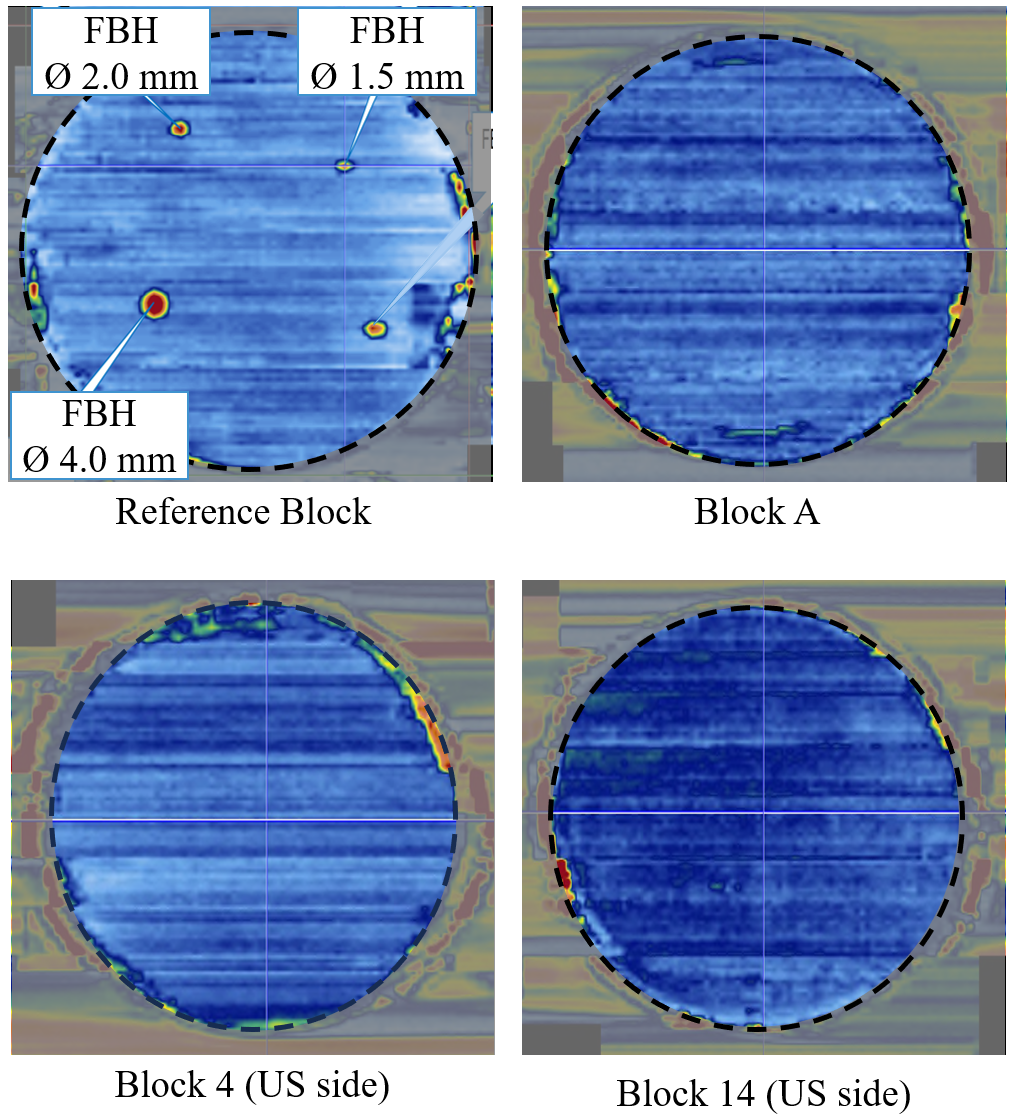}
\caption{Ultrasonic C-scan results, in which discontinuity is shown in red and continuity is shown in blue. FBH: flat bottom hole.}
\label{fig:UT_Result}
\end{figure}

\subsubsection{Ultrasonic testing}
All bonding interfaces (US and DS) of all (un)irradiated blocks were inspected by UT (Fig.~\ref{fig:UT_Result}). No indications of debonding---such as cracks parallel to the surface, delamination, or volumetric flaws---were detected at the interfaces or in their vicinities. Some blocks exhibited an increased echo close to the interface depth; however, this was homogeneous over the blocks' surfaces without deflections, and it was far below the detection threshold for discontinuity. Different echo levels were observed mostly at W interfaces without evident correlation to their radiation levels.

\begin{figure}[htbp]
\centering
\includegraphics[width=0.9\columnwidth]{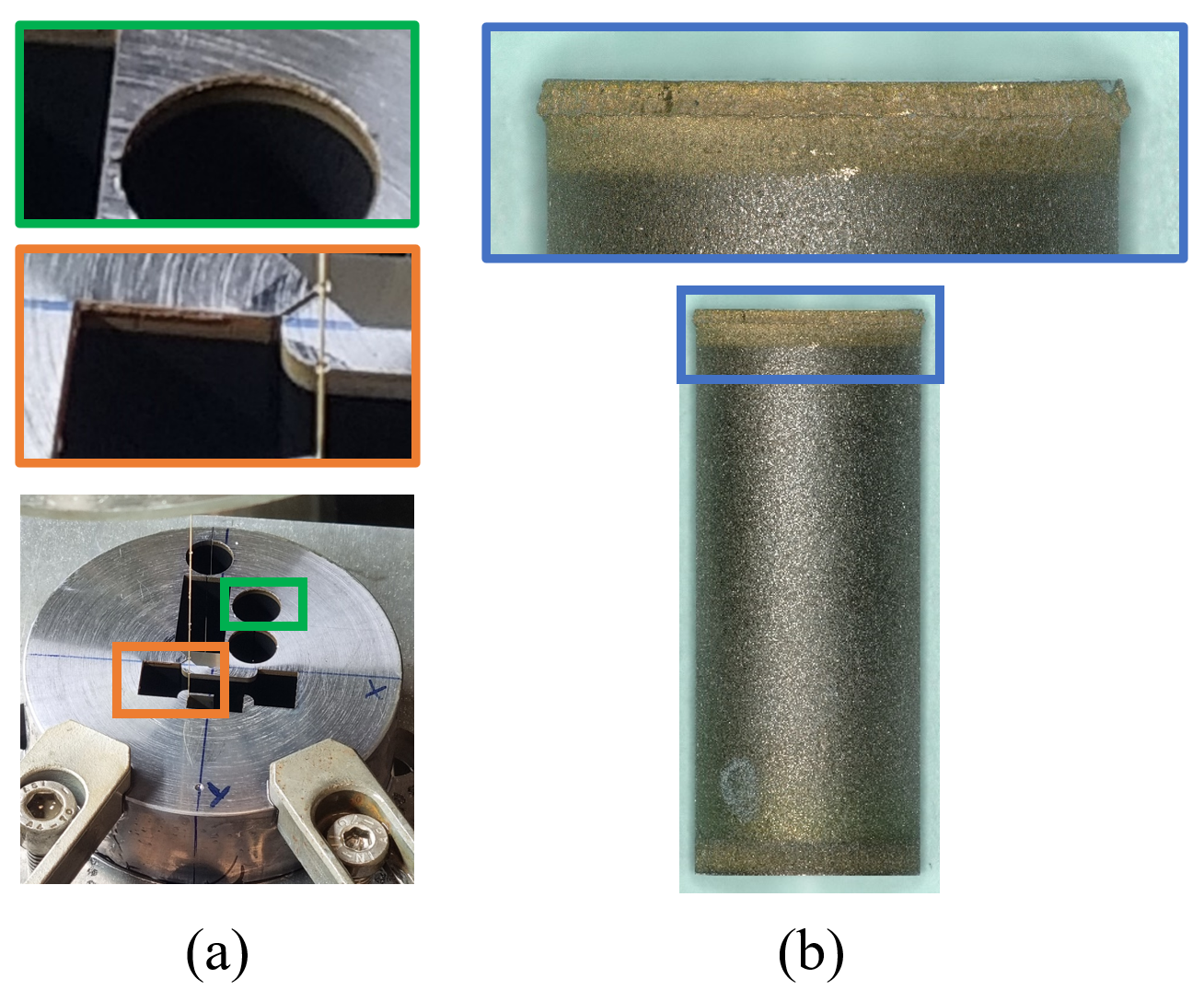}
\caption{Burrs detected on the upper side of (a)~Block~A and (b)~the thermal-diffusivity cylinder after the first cutting step.}
\label{fig:burr}
\end{figure}

\subsection{Destructive tests}
\subsubsection{Specimen extraction}
During the specimen extraction, the EDM wire consistently fractured and needed to be re-fed continuously during the first separation cuts from the blocks. The morphology of the W surface after EDM cutting showed intergranular separations and areas with detached W grains. In addition, the tensile specimen 14T3 and the shear specimen 14S2 fractured during their final EDM cutting steps before testing. The fracture of 14T3 appeared close to the interface on the DS side. During the first cutting steps, deformations inside the cladding bulk material were detected (see Fig.~\ref{fig:burr}). This behavior was only observed sporadically on the upper block sides that had first contact with the water jet.

\begin{figure}[htbp]
\centering
\includegraphics[width=\columnwidth]{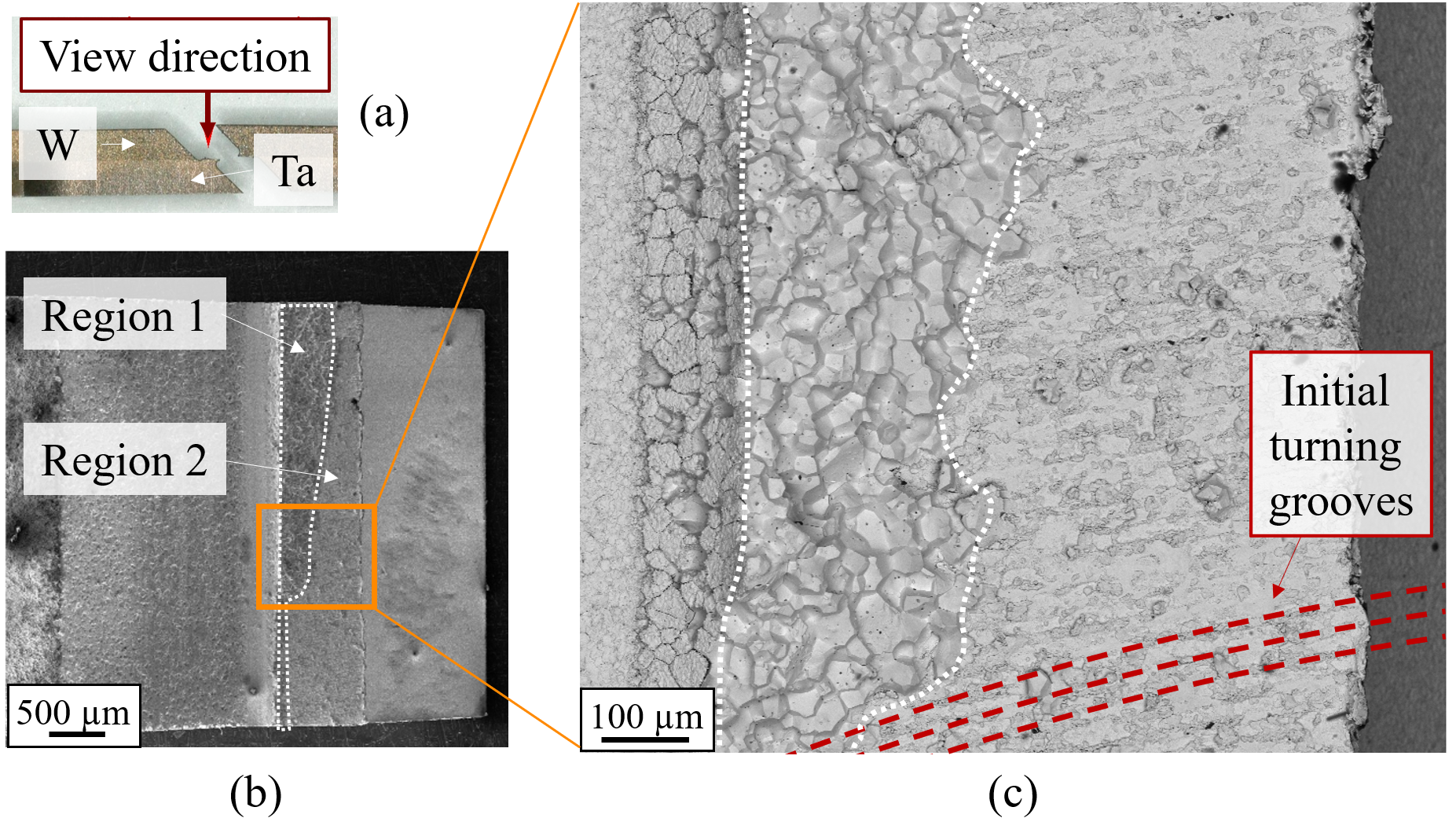}
\caption{Examination of 14S2, which fractured during the final cutting step, revealed an intergranular fracture at the interface in Region~1; in contrast, Region~2 showed no distinct fracture characteristics.}
\label{fig:14S2}
\end{figure}

A comprehensive fractography examination of 14S2 was performed by analyzing the Ta cladding side using SEM (see Fig.~\ref{fig:14S2}). The fracture occurred along the interface, and two fracture regions (regions~1 and 2) were distinguished. SEM and EDS assessed Region~1 as comprising intergranular fractures through the W material, appearing parallel to the interface, and with sporadic cleavage-fracture characteristics; the visible homogeneous grain size matched the grain size found in the bulk W material. Region~2 exhibited interleaved bands with different fracture characteristics, and these bands showed a curvature attributable to concentric circles. At the microscopic level, multiple fracture modes were detected. Most frequently encountered were areas with no distinct fracture characteristics, in which the fracture-initiation site and propagation direction could not be determined.

\begin{figure*}[htbp]
\centering
\includegraphics[width=2\columnwidth]{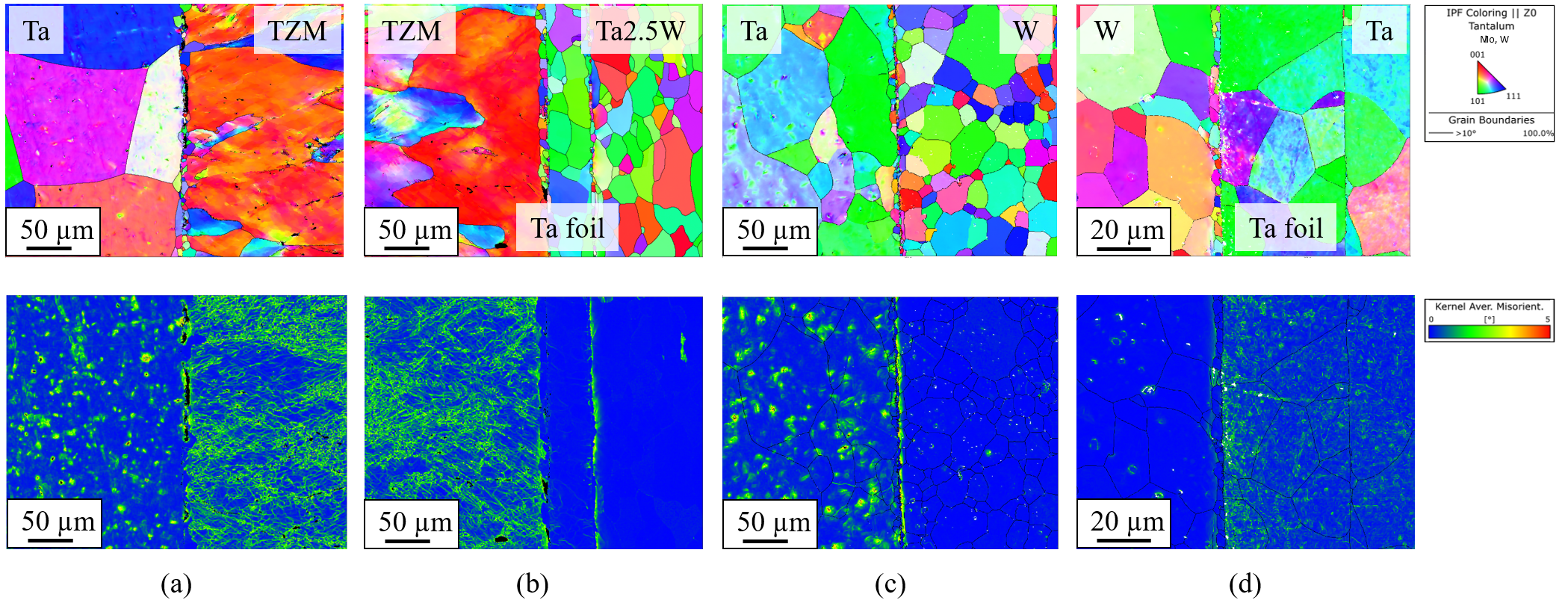}
\caption{EBSD inverse pole figures (IPFs) in $z$ direction and local misorientation maps at the interfaces of (a)~Block~A, (b)~Block~4 (DS side), (c)~Block~B, and (d)~Block~14 (DS side).}
\label{fig:EBSD_all}
\end{figure*}

\begin{figure}[htbp]
\centering
\includegraphics[width=\columnwidth]{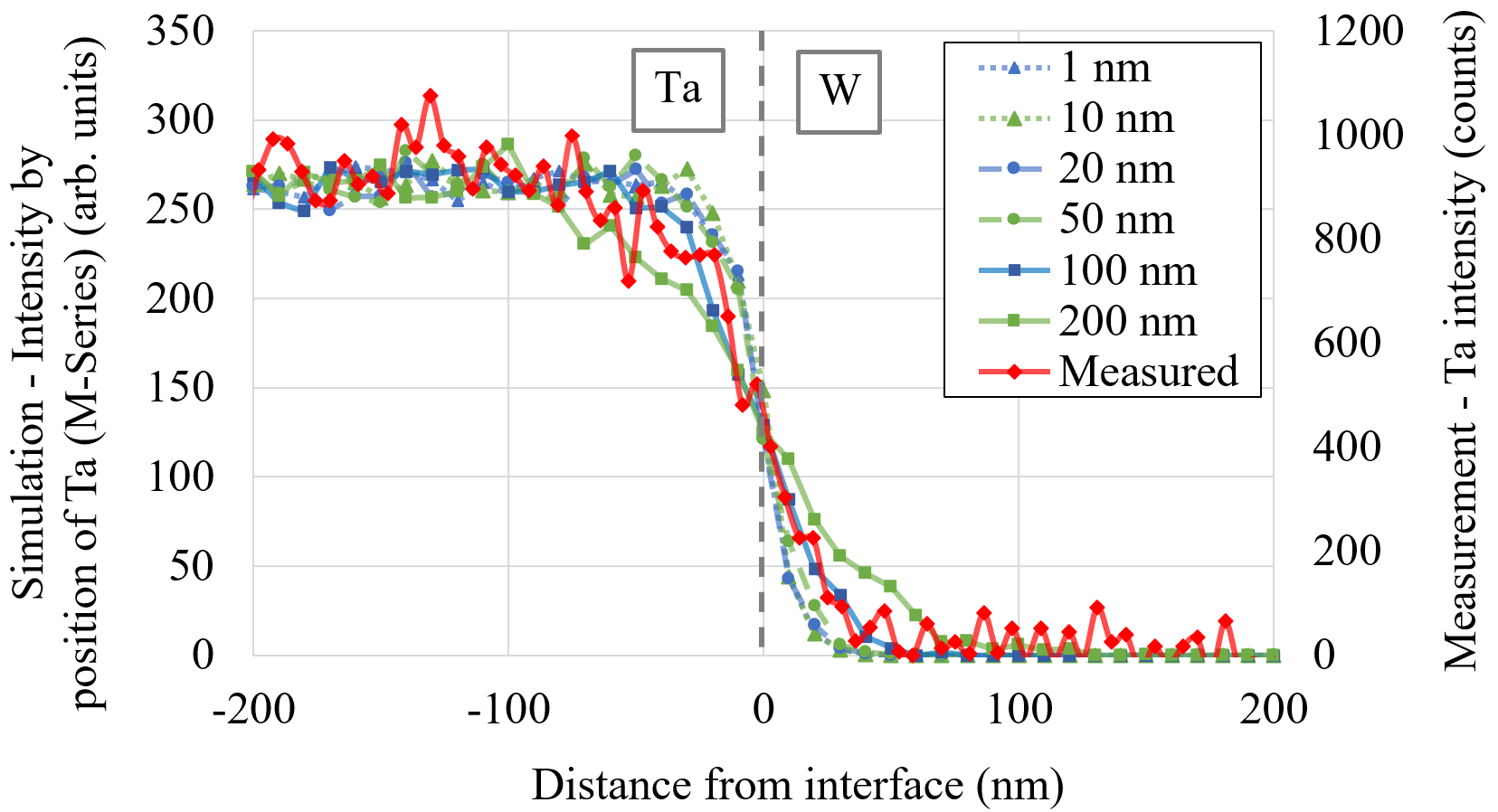}
\caption{Ta signal perpendicular to the interface measured by EDS mapping compared with the simulated Ta signals from electron beams of different diameters.}
{\justifying
}
\label{fig:interface_bonding}
\end{figure}

\subsubsection{Microstructural examination}
Examinations of the microstructure of specimen AM1 revealed that etching led to the appearance of gaps in the Ta/TZM interface caused by imperfect adhesion between the parent materials. Although preparation without etching yielded a gap-free interface, the established procedure could not guarantee unaffected areas, which implies that regions with loose adhesion existed before extraction, and they expanded during sample preparation.

Figure~\ref{fig:EBSD_all} shows EBSD and local misorientation maps close to the interfaces. The TZM core material presented elongated grains oriented in the axial direction of the blocks and an intragrain area with a recovered microstructure, which is characteristic of stress-relieving heat treatment; no recrystallization was observed. The subgrains were well distinguishable within the parent grains. Different kinds of second-phase particles, nonmetallic inclusions, and precipitates were observed; specifically, second-phase particles such as oxidic nonmetallic inclusions, Ti carbides, and complex Zr/Ti carbides were visible. The microstructure near the interface was equivalent to that of the core. The W core materials exhibited a fully recrystallized microstructure featuring inter- and intragranular porosity. Using SEM, the porosity was determined to be 2.2\%. Pure Ta exhibited the largest grain size, and this was followed by the Ta foil and Ta2.5W, which presented a smaller grain size.

In the vicinity of the interfaces, all materials, except for the Ta-foil/Ta interface, exhibited recrystallized fine grains. For example, the TZM/Ta-foil interface showed a single layer of recrystallized grains on the TZM side, approximately \SI{10}{}-\unit{\micro\meter} thick (see Fig.~\ref{fig:EBSD_all}). These grains were fully recrystallized and slightly elongated along the interface with the adjacent nonrecrystallized area, displaying a diminishing strain gradient toward the unaffected recovered core material. Similarly, the other
interfaces featured a narrow band of recrystallized grains followed by a narrow strained region. In contrast, the Ta-foil/Ta interface of 14M2 did not present new-formed grains or strain gradients. The Ta foils were completely recrystallized without any strain gradients.

The diffusion bonding of BM1 (Ta/W) was determined by EDS mapping, which showed multiple Ta signals on the W side in the form of tips. The longest tips had a depth of \SI{180}{nm} from the interface, while along the interface, the tips occurred at distances of 0.1--\SI{1.0}{\micro\meter}, similar to the size of the newly formed W grains at the interface. In addition, an EDS line scan was performed perpendicular to the interface, which revealed an S-shaped logistic function. Comparing the measured EDS signal of Ta with the simulation results showed a correlation with the 100-nm beam diameter (see Fig.~\ref{fig:interface_bonding}). The expected electron-beam diameter of the machinery used was \SI{10}{nm}, meaning that the beam diameter was ten times that predicted.

For the TZM, the KAM evaluations of AM1 and 4M1 showed significant numbers of low-angle grain boundaries without qualitative or quantitative differences. No increased strain due to beam-induced thermal stresses was identified, nor were there any signs of decreased strain, progressed recovery, or recrystallization due to increased temperatures during irradiation. The KAM evaluation of the W core material indicated minimal strain in the matrix, and areas with high local misorientation were caused by microsection surface-preparation artifacts at pore locations. In conclusion, the unirradiated and irradiated specimens were considered to be equivalent.

\begin{figure}[htbp]
\centering
\includegraphics[width=0.7\columnwidth]{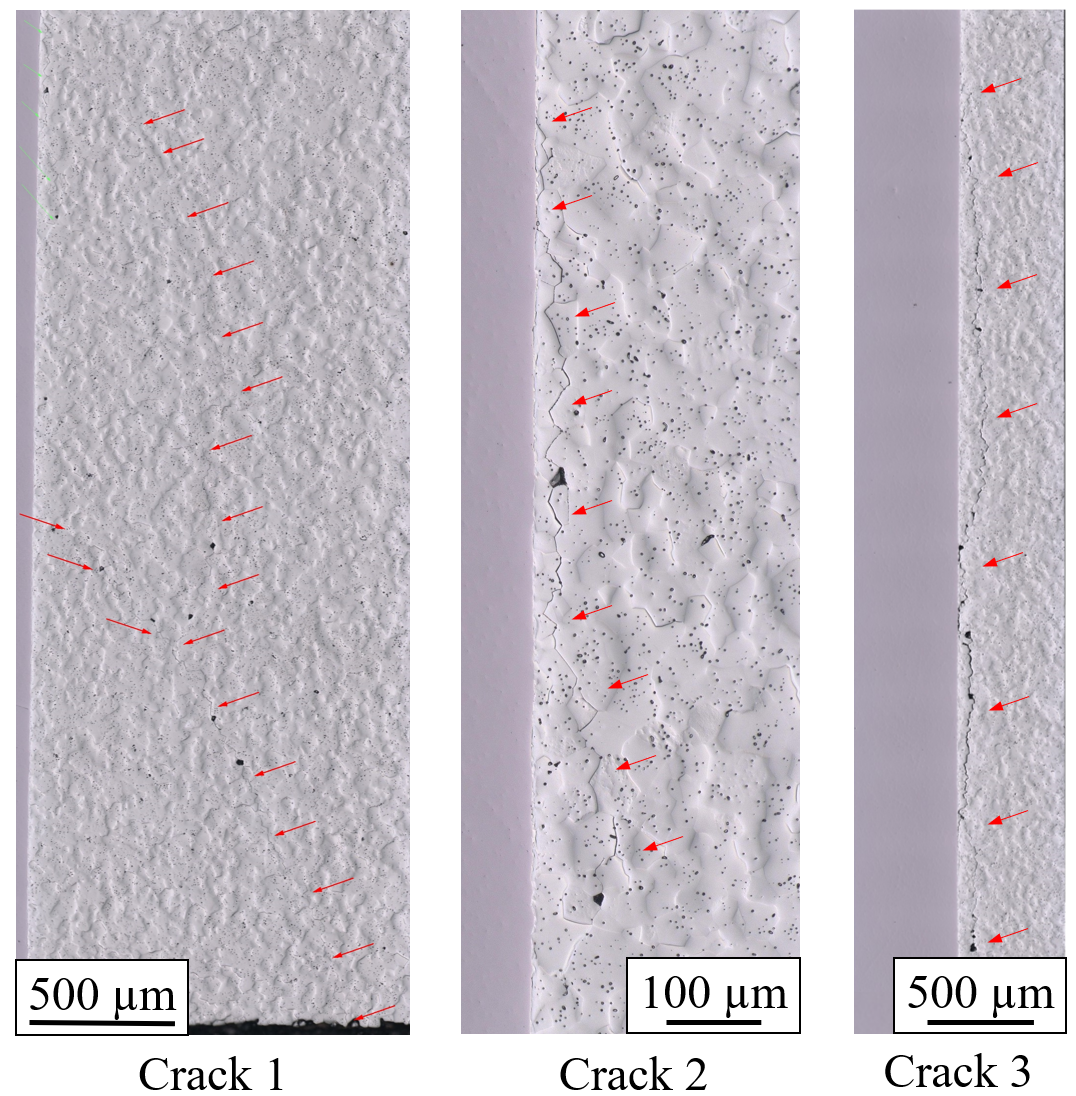}
\caption{Light optical microscopy of 14M1 (W) showing the three cracks detected in the W.}
\label{fig:14M1_cracks}
\end{figure}

\begin{figure}[htbp]
\centering
\includegraphics[width=\columnwidth]{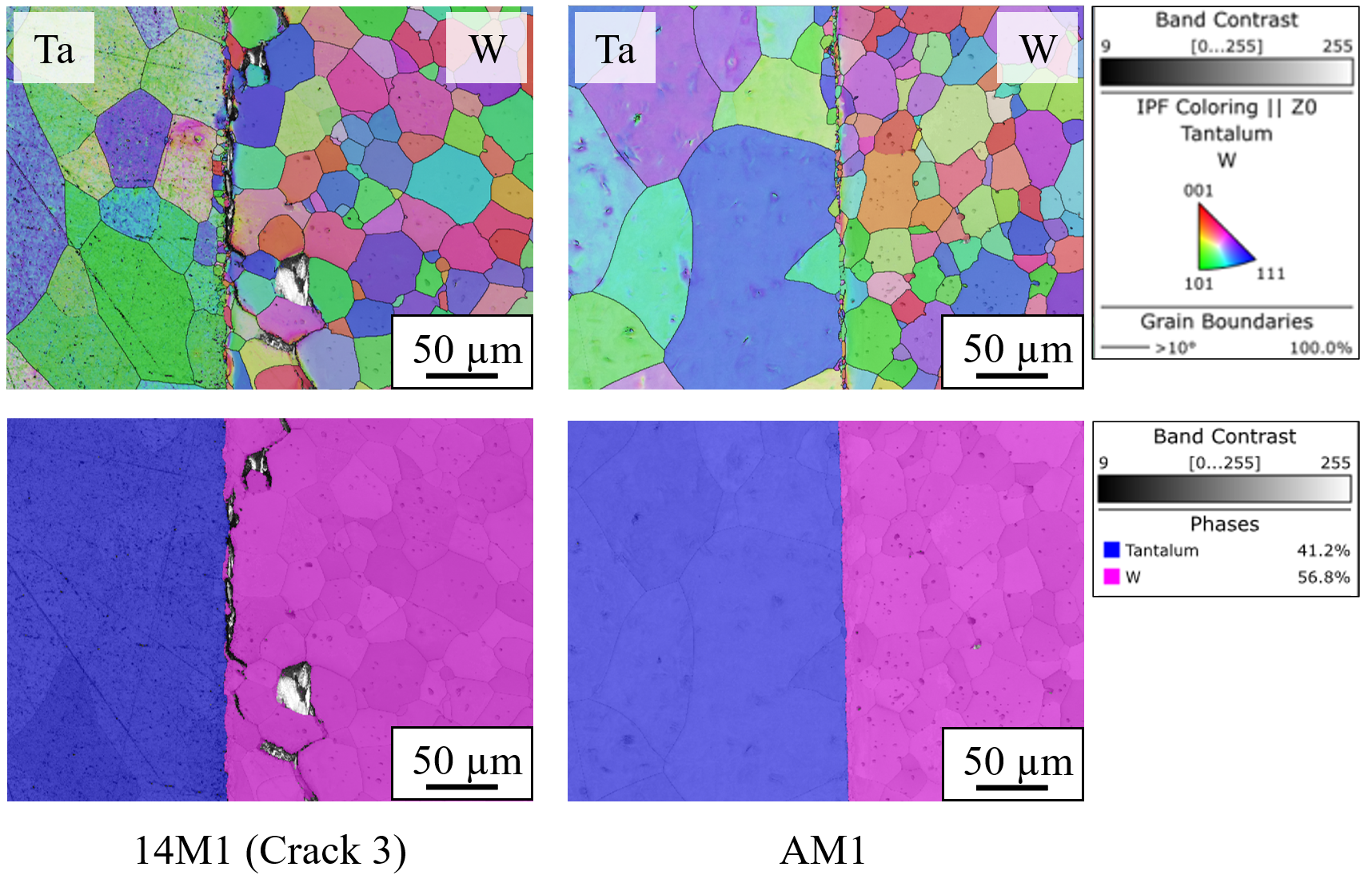}
\caption{EBSD inverse pole figures (IPFs) and phase maps at the interfaces of Block~14 (US side), and Block~B.}
\label{fig:EBSD_2}
\end{figure}

Several cracks were revealed in 14M1 at the US Ta/W interface (see Fig.~\ref{fig:14M1_cracks}). Two cracks, cracks~1 and 2, were located close to the beam-impact axis, and Crack~3 was detected approximately \SI{16}{mm} from the beam-impact axis. Crack~1 had a length of approximately \SI{3}{mm} parallel to the interface and extended approximately \SI{1}{mm} into the core. Cracks~2 and 3 propagated along the interface on the core side, with lengths of 0.8 and \SI{4.3}{mm}, respectively. Figure~\ref{fig:EBSD_2} displays EBSD and phase images of Crack~3, showcasing the fracture in the W core material. The crack propagation was intergranular, in the W core and directly at the interface; hence, these cracks were not considered to be detachments of the interface.

\begin{figure}[htbp]
\centering
\includegraphics[width=\columnwidth]{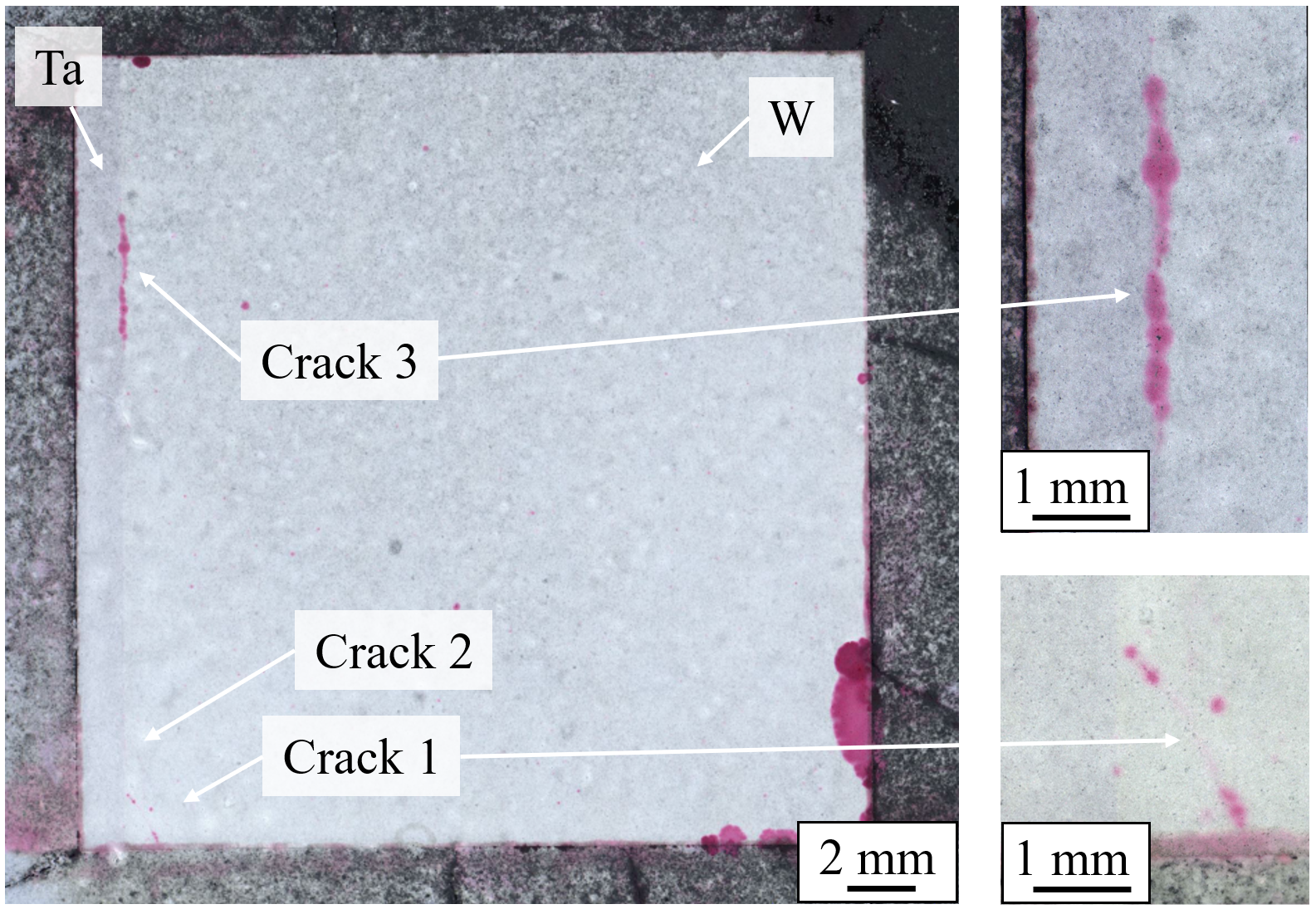}
\caption{Sample 14M1 after dye penetrant testing, which did not indicate Crack~2.}

\label{fig:PT_results_cracks_14}
\end{figure}

\begin{figure}[htbp]
\centering
\includegraphics[width=\columnwidth]{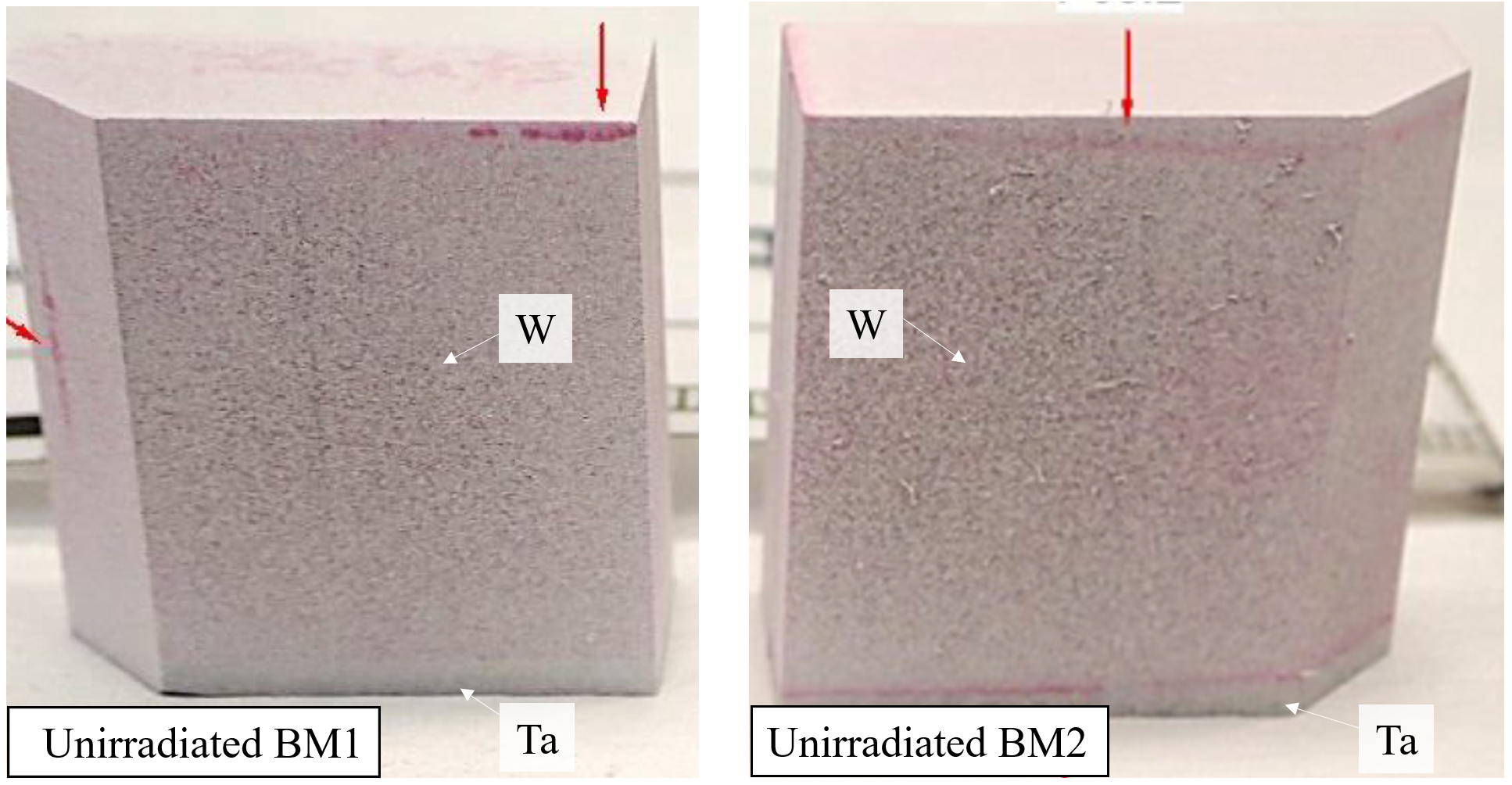}
\caption{Dye penetrant testing performed on the unirradiated samples BM1 and BM2.}
\label{fig:PT_results_cracks_19}
\end{figure}

\begin{figure}[htbp]
\centering
\includegraphics[width=\columnwidth]{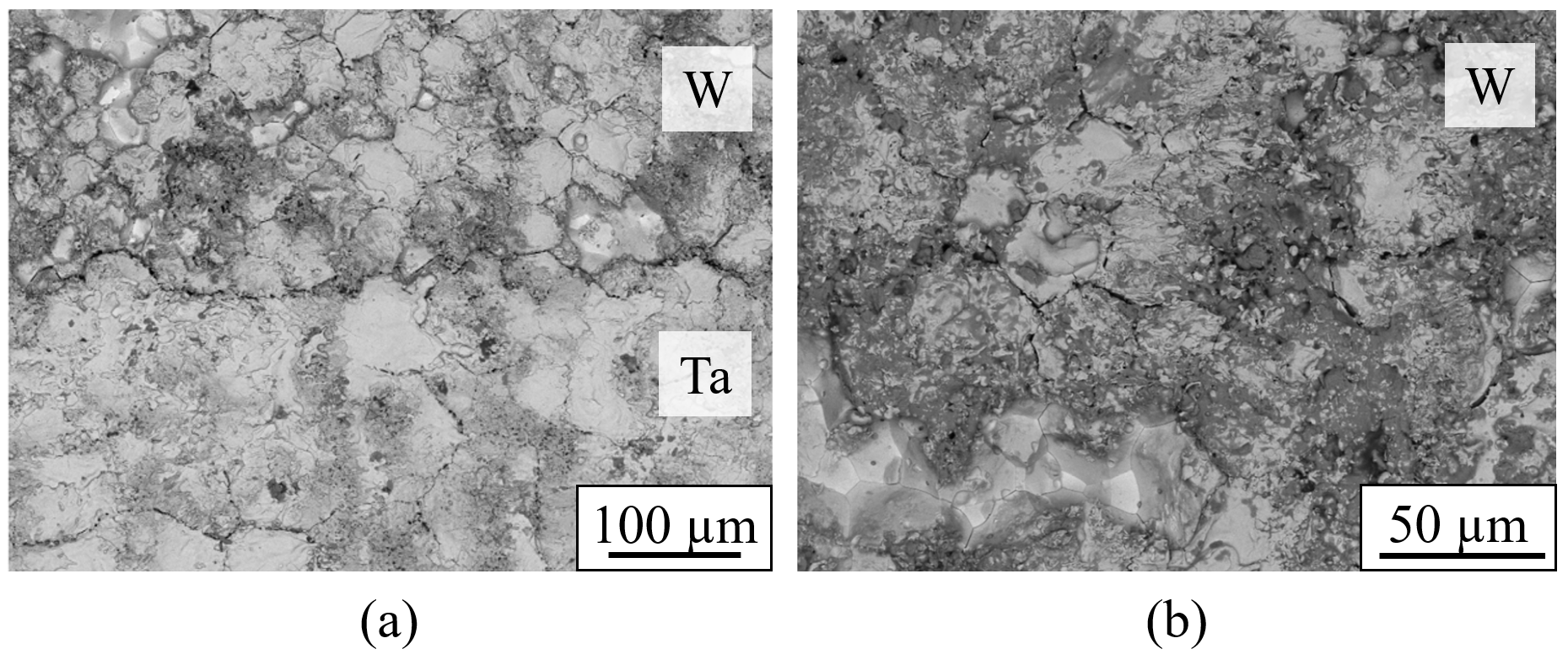}
\caption{SEM images of cracks found in the unpolished specimen BM2 (a)~at the interface and (b)~in the core close to a cutting edge.}
\label{fig:PT_19M2}
\end{figure}

Dye PT was performed on 14M1, detecting cracks~1 and 3; Crack~1 was only partially reproduced, and Crack~2 was not identified at all (see Fig.~\ref{fig:PT_results_cracks_14}). Figure~\ref{fig:PT_results_cracks_19} shows the dye PT results for BM1 and BM2; a reddish color was present in the background due to the unpolished surface. SEM images were taken at the PT-indicated locations, and it was found that BM1 featured grinding marks, detached W grains, and intergranular separations in the W bulk material. Specimen BM2 showed several linear diffuse indications in the area of the interface, while none were visible in BM1. Figure~\ref{fig:PT_19M2} presents the unpolished surface of BM2 at the interface and at the core close to the cutting edge. Both areas of the W surface exhibited similar characteristics, such as intergranular cracks and detached W grains, while the majority of the surface was covered by a recast layer due to the EDM. Considering the penetration and dwell time, the coloration indicated that all the cracks observed in 14M1, BM1, and BM2 were superficial.

\begin{table}[htbp]
   \centering
   \footnotesize
   \renewcommand{\arraystretch}{1.2}
   \setlength{\tabcolsep}{2pt}
   \caption{Comparison of the measured Vickers hardness of the unirradiated (AM1, BM1, and BM2) and irradiated (4M1 and 14M1) samples (HV0.5) with small-scaled Ta-alloy-clad prototype capsules (HV5) and comparable literature values.}
   \begin{threeparttable}
   \begin{tabular}{l*{8}{c}}
       \toprule
       \toprule
            &&      Literature &&  Prototype && Unirrad. && Irrad. \\
              &&       &&  capsules &&   &&   \\
        \midrule
           TZM     &&   235--250\tnote{a}  && 240\tnote{e}, 215\tnote{f}  && $260 \pm 10$   && $257 \pm 13$ \\
           W       &&   350\tnote{b}        && 435\tnote{e}, 355\tnote{f}  && $346 \pm 10$   && $356 \pm 10$ \\
           Ta      &&   75--105\tnote{c}   && 80\tnote{e}, 70\tnote{f}   && $88 \pm 6$     && $96 \pm 5$\\
           Ta2.5W  &&   160--240\tnote{d} && 140\tnote{e}, 120\tnote{f}  && --             && $161 \pm 7$\\
       \bottomrule
       \bottomrule
    \end{tabular}
    \vspace{1mm}
    \begin{tablenotes}
        \footnotesize
        \item [a]Annealed TZM~\cite{ghazali2020mechanical}.
        \item [b]Fully recrystallized (annealed?) W~\cite{tanure2019effect, alfonso2014recrystallization, alfonso2015thermal, serret2011mechanical}.
        \item [c]Annealed Ta~\cite{clark1991effect, jones1974alamos}.
        \item [d]Annealed Ta3W~\cite{alaei2019development, jones1974alamos}.
        \item [e]After HIPing at 1200\,\unit{\degreeCelsius} for \SI{3}{h}~\cite{busom2020application}.
        \item [f]After HIPing at 1400\,\unit{\degreeCelsius} for \SI{3}{h}~\cite{busom2020application}.

    \end{tablenotes}
    \end{threeparttable}
   \label{tab:hardness_results}
\end{table}

\subsubsection{Indentation hardness testing}
The indentations were performed along three profiles without showing dependencies regarding the different distances to the beam axis. Therefore, the Vickers hardness results were considered for each bulk material and irradiation state individually, and it was considered annealed because the blocks of the prototype target were HIPed at 1200\,\unit{\degreeCelsius} for \SI{3}{h}. Both of the unirradiated blocks were clad with Ta; hence, no results for the unirradiated Ta2.5W were obtained. Table~\ref{tab:hardness_results} lists these results alongside comparable values from the literature and from previously manufactured and tested prototype capsules. The Vickers hardness of the irradiated and unirradiated samples showed no differences, and they correlated with the expected results.

\begin{table}[htbp]
   \centering
   \footnotesize
   \renewcommand{\arraystretch}{1.2}
   \setlength{\tabcolsep}{6pt}
   \caption{Yield strength (YS) and ultimate tensile strength (UTS) of the unirradiated AT and irradiated 4T and 14T tensile specimens (all values in MPa).}  \begin{tabular}{l*{8}{c}}
       \toprule
       \toprule
       & \multicolumn{2}{c}{TZM (AT)} &&\multicolumn{2}{c}{TZM (4T)} && \multicolumn{2}{c}{W (14T)} \\ \cline{2-3}  \cline{5-6} \cline{8-9}
            & YS & UTS &&  YS & UTS &&  YS & UTS \\
        \midrule
           22\,\unit{\degreeCelsius}                  & \SI{552}{} & \SI{569}{} && \SI{511}{} & \SI{535}{}    &&  --       & \SI{84}{}  \\
           200\,\unit{\degreeCelsius} & \SI{376}{} &\SI{448}{}  && \SI{343}{} &\SI{409}{}     && \SI{288}{} &\SI{334}{} \\
       \bottomrule
       \bottomrule
    \end{tabular}
   \label{tab:mechanical_results}
\end{table}

\begin{figure*}[htbp]
\centering
\includegraphics[width=2\columnwidth]{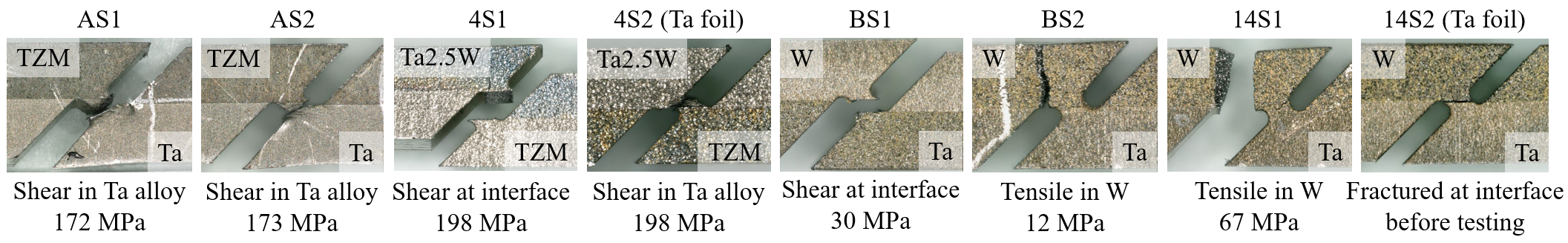}
\caption{Fractures of the shear specimens after testing with each fracture type, position, and breaking strength at 22\,\unit{\degreeCelsius}.}
\label{fig:shear}
\end{figure*}

\subsubsection{Mechanical testing}
Table~\ref{tab:mechanical_results} lists the yield strength (YS) and UTS values obtained from the tensile tests. Comparing the TZM results, it was found that the YS and UTS values of the irradiated 4T sample were 30--\SI{40}{MPa} lower for both temperatures. The W bulk material of Block~14 exhibited premature failure; the YS and UTS were not reached at 22\,\unit{\degreeCelsius} and the UTS was not achieved at 200\,\unit{\degreeCelsius}. However, W showed reproducibility in terms of stress and elongation at break for both temperatures and for the YS at 200\,\unit{\degreeCelsius}.

During the shear-specimen testing, different types of fracture were observed, such as tensile fractures in the W and shear fractures in the cladding material or at the interface (see Fig.~\ref{fig:shear}). The reported stresses were calculated with an interface area of $1.0 \times 3.6$~mm$^2$ for each specimen. For the TZM core materials, the extension to fracture was in the range 80--\SI{750}{\micro\meter}, with the unirradiated samples (AS1 and AS2) failing at higher strain values than the irradiated ones (4S1 and 4S2). Sample 4S1 showed the lowest overall strain and presented the only shear fracture at the interface, while the other three [AS1, AS2, and 4S2 (Ta foil)] failed in the Ta alloy close to the interface. All W shear specimens exhibited very low extension to their fracture values (\SI{<10}{\micro\meter}). Two specimens (BS2 and 14S1) displayed tensile fractures in the W bulk material far away from the interface region: BS1 had a shear fracture at the interface, and 14S2 fractured at the interface during the final EDM cutting step (before testing).

\subsubsection{Thermal testing}
The density measurements for pure W yielded a value of \SI{18.720}{g\,cm^{-3}}, indicating a porosity of 3.0\%. In the 4D (Ta2.5W/TZM) specimens, the bonding interfaces were approximately in the middle of the thickness, whereas the cladding layers of the 14D (Ta/W) samples were significantly thicker than the core, particularly in the even-numbered samples (those with Ta foil). The thermal contact resistances ($R$-values) of all odd-numbered samples (without a Ta interlayer) were negligible, while the values for the even-numbered 4D samples ranged from \SI{1e-7}{} to \SI{25e-7}{m^2\,K\,W^{-1}}, and the values for the 14D samples ranged from \SI{1e-7}{} to \SI{10e-7}{m^2\,K\,W^{-1}}. No temperature dependency, influences from the Ta interlayer or HIPing parameters, or beam-induced effects were observed. The TCC analysis determined an increase of 10\,\unit{\degreeCelsius} in the maximum temperatures with a TCC of \SI{4e4}{m^2\,K\,W^{-1}} at the bonding interfaces, corresponding to an $R$-value of \SI{250e-7}{W\,m^{-2}\,K^{-1}} for both Ta2.5W-clad blocks~4 and 14; hence, the $R$-values of the bonding interfaces were negligible.

\subsection{Temperature instrumentation}

\begin{figure}[tbp]
\centering
\includegraphics[width=\columnwidth]{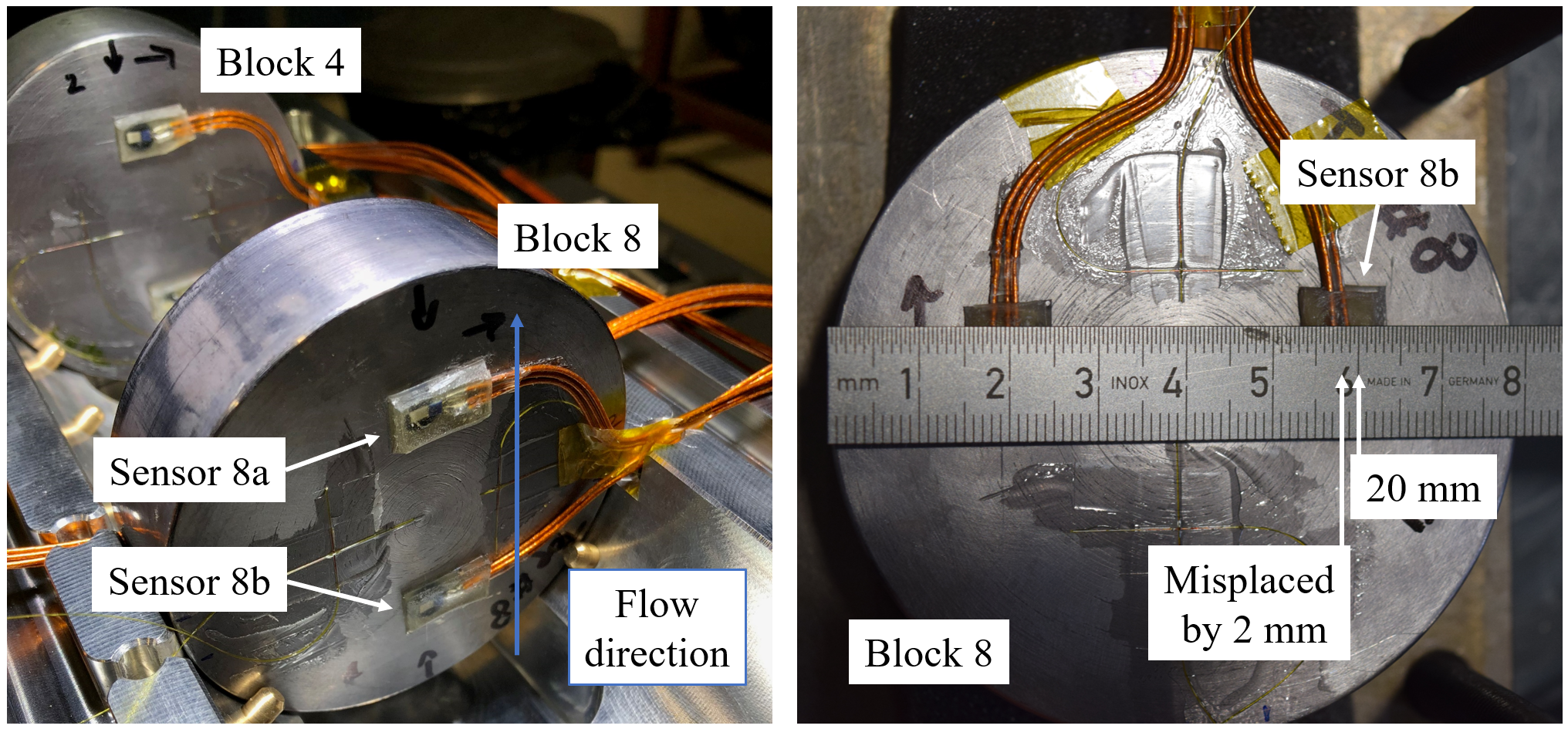}
\caption{Temperature instrumentation installed on Block~8 before irradiation.}
\label{fig:sensor}
\end{figure}

During the operation of the prototype BDF target, increased temperatures were measured by one of the sensors attached to Block~8. The PIE did not result in any findings for Block~8 that could explain the behavior of Sensor~8b (see Fig.~\ref{fig:sensor}); however, upon examination of past data from the target instrumentation, some observations stand out. Sensor~8b was placed in the only water channel that had temperature sensors on both sides (US and DS). In general, the water channels had a width of only \SI{5}{mm}, and the temperature sensor on the opposite side to Sensor~8b failed during operation. The flow of the cooling water started from Block~1 and passed sensors~8b and 9b of blocks~8 and 9 first; this could have resulted in a small amount of turbulence due to the sudden change of flow. The sensor was constantly attached to the block, because no noise was measured and its temperature reading showed an appropriate reaction to the pulse-cycle time. It was determined that no cavities were present in the cover agent because the base temperature was not increasing. The heating ramps observed during impacts correlated with those from the other three sensors in the same water channel, but the cooling rate measured by Sensor~8b did not match those of the other sensors.

Thermal simulations were conducted using \textsc{ansys} {\footnotesize Mechanical} to assess the temperature sensitivity by varying the cooling factor and the distance between the sensor and the beam impact. Both simulations showed that the temperatures recorded by Sensor~8b could be reached if, for example, the cover agent of the sensor did not allow any cooling effect from the water (a heat-transfer coefficient of zero) or if the sensor was at a distance of \SI{12}{mm} from the beam impact; however, the sensor was misplaced by a maximum of \SI{2}{mm} toward the beam-impact point, which means that it was \SI{16}{mm} away considering the beam offset of \SI{2}{mm}.

\FloatBarrier
\section{Evaluation of test results}
\subsection{Location of beam impact}
Based on the results from the Gafchromic film and the BTV system, the location of the beam impact was identified to be in the lower-right corner of the target, and the two derived locations were separated by vertical and horizontal distances of approximately \SI{1}{mm}. The beam positions on blocks~4 and 14 were defined for each by the average of the US and DS sides' Gafchromic film results, and this resulted in beam-impact offsets of (1.5~mm, $-$0.7~mm) and (2.3~mm, $-$0.2~mm), respectively.

\subsection{Bulk material}
\subsubsection{TZM}
The YS and UTS values of TZM from the same material batch were previously determined using a strain rate of \SI{9e-3}{s^{-1}}~\cite{FHinternal2017}. In the present study, a lower strain rate of \SI{2.5e-4}{s^{-1}} was used to simulate quasistatic testing conditions. The YS and UTS values of the unirradiated Block~A were found to be 50--\SI{100}{MPa} lower than the previous measurements.
When comparing the YS and UTS of TZM before and after irradiation, a degradation of 30--\SI{40}{MPa} was observed. Due to machine failure, specimens AT and 4T had to be examined using different testing machines; however, the calibration documents, load-train setup, strain rate, and loading rate did not indicate any shortcomings. Calculating the YS/UTS ratio, AT and 4T presented values of \SI{0.97}{} and \SI{0.95}{} at 22\,\unit{\degreeCelsius} and \SI{0.84}{} and \SI{0.84}{} at 200\,\unit{\degreeCelsius}, respectively. Hence, the degradation seemed systematic. During operation, the material was subjected to thermal loads up to 200\,\unit{\degreeCelsius}. Irradiation damage was excluded because the PoT value of \SI{2.4e16}{} was insignificant, and an increase in mechanical properties would be expected due to irradiation hardening. The TZM microstructures of AM1 and 4M1 presented no differences, although the examination was performed on localized areas and is hence not representative of the bulk material. Therefore, the cause of the degradation cannot be determined because no mistakes were detected in the testing setup, thermal relaxation of the TZM seemed impossible considering that the recrystallization temperature is above 1000\,\unit{\degreeCelsius}, and similar hardness values were measured.

\subsubsection{Tungsten}
The sintered W exhibited porosity values in the range 2.2\%--3.0\% and a fully recrystallized microstructure, which leads to brittle behavior and low mechanical properties~\cite{alfonso2014recrystallization}. Because EDM damages the W microstructure at the surface level, different cutting methods should be chosen when undertaking mechanical material characterization.

Multiple factors were taken into consideration when attempting to establish the origin of the W cracks in 14M1 (US). During the examination of 14S2, intergranular separations, areas with detached W grains, and a recast layer were observed along the cutting plane induced by the EDM. A recast layer can develop tension stresses, which can support the development of hot or cold cracks. In addition, the sample exhibited high brittleness originating from a fully recrystallized grain structure and high porosity resulting from sintering. For example, the tensile specimen 14T3 broke during the cutting process on the DS side within the W material. While the exact levels of residual stresses in the materials were unknown, very high values---especially close to the interface---were expected according to the results of simplified simulations, the large differences in the thermal-expansion coefficients of Ta and W (\SI{6.1E-06}{} and \SI{4.1E-06}{K^{-1}} at 25\,\unit{\degreeCelsius}, respectively~\cite{FHinternal2017}), and the continuous breakage of the cutting wire during specimen extraction. Dye PT of 14M1 identified superficial cracks, while similar cracks were found in the unirradiated BM2 sample close to the interface. In addition, one crack was observed in the center close to the final separation cut that had propagated through both samples (BM1 and BM2). Nonetheless, based on the results of the simplified simulations, low residual stresses were expected in the center after HIPing. Additionally, the results of UT of all flat surfaces revealed neither the cracks in 14M1 nor those in BM2 that were parallel to the interface. It is suspected that the combination of high residual stresses, brittleness, damage to the grain structure during EDM cutting, and potentially beam-induced thermal stresses contributed to the development of these cracks. Hence, based on the findings from the unirradiated samples, the cracks observed in 14M1 were not considered to be caused by the beam.

\subsubsection{Ta alloys}
No relevant geometrical changes such as swelling, corrosion, cracks, separations, pores, or melting were observed in the Ta alloys; therefore, it was considered that both cladding materials---Ta and Ta2.5W---were reliable.
The flat surfaces in the center region exhibited Ra values above the expected threshold and strongly pronounced turning grooves; this led to a higher level of carbon deposition in combination with the beam impact. It was suspected that this was caused by insufficient lubrication during the turning process after HIPing. In addition, lubricant residue was observed, which can cause local heating and increase the carbon concentration. As carbon deposition on the target blocks is undesirable, it is recommended that the surface roughness in the center should be decreased and the cleaning process of the blocks should be improved.

Comparing Ta and Ta2.5W, both exhibited similar results, and no debonding or defects in the cladding materials were observed. A previous study~\cite{busom2020application} determined that the interface bonding strength was higher with Ta2.5W for both core materials. In addition, both Ta alloys demonstrated negligible $R$-values, and the thermal conductivity of the bulk materials is slightly higher for pure Ta than for Ta2.5W. To guarantee a long target lifetime, plastic deformation should be avoided; as such, simulations were performed, and the Christensen failure criterion was applied for both cladding materials. Under the operational temperature of 200\,\unit{\degreeCelsius}, the yield strengths of Ta and Ta2.5W are 121 and \SI{227}{MPa}, respectively (see Table~\ref{tab:FH_mechanical_results}); therefore, Ta2.5W is recommended as the cladding material for the BDF target.

\subsection{Bonding interfaces}
\subsubsection{Formation of new grains}
Narrow recrystallized bands with fine grains were observed at the bonding interfaces (see Fig.~\ref{fig:EBSD_all}). It is assumed that plastic deformation was induced in the near-surface region during machining; after this, HIPing with elevated temperature and pressure (1200\,\unit{\degreeCelsius} and \SI{150}{MPa}) initiated recrystallization near the surface region. In addition, during the cool-down phase of the HIPing process, the interfaces were exposed to particularly high stresses, and this aided the formation of recrystallized bands because increased stresses lower the recrystallization temperature. In general, their presence has no negative effect on the quality of the bonding interfaces.

The recrystallization bands were present on all interfaces except for the Ta-foil/Ta interface in 14M2. The Ta-foil interfaces generally behaved differently because the bands were only detected on the core-material side (TZM and W). Fine grains were visible in the Ta-foil/Ta2.5W interface of 4M1, with a strain gradient visible toward the Ta2.5W side; however, it could not be determined whether these grains belonged only to Ta2.5W or also to the Ta foil. In addition, the Ta-foil/Ta interface of 14M2 did not exhibit any fine grains. One explanation for this could be the presence of lower stresses along the interface, as both materials have the same coefficient of thermal expansion. It seems that the Ta foils did not form fine grains during the HIPing process, and this could be due to the different manufacturing procedure. Overall, no differences were detected between the irradiated and unirradiated interfaces.

\subsubsection{Signs of bad bonding quality}
Surface preparation without an etchant resulted in gap-free interfaces; however, without etchant, the surface quality was inadequate for obtaining clear channeling contrast or high-quality EBSD images. When using interim etching and etch-polishing, gaps appeared at the interfaces, and the duration and aggressiveness of the etchant correlated with their length and width. As the interface was attacked more strongly by the etchant than the surrounding materials, there must have been preexisting narrow gaps; therefore, it was suspected that the pre-etchant gap-free interfaces were caused by smearing of the ductile Ta into the gaps during polishing. It was concluded that there was nonperfect adhesion between the cladding and core materials; nevertheless, this behavior was observed in both the unirradiated and irradiated interfaces, thus indicating that any deterioration of the bonding interfaces resulting from beam-induced thermomechanical stresses was immeasurable.

Indications of inadequate diffusion bonding were also observed in the fractured shear specimen 14S2 (see Fig.~\ref{fig:14S2}). Region~2 exhibited concentric circles, which can be interpreted as the initial turning grooves from the Ta or W blocks; hence, it was concluded that a decohesion fracture was initiated in Region~2 and progressed toward Region~1.

\subsubsection{Signs of good bonding quality}
The shear specimens that did not fracture at the interface presented higher interface strengths than the fractured bulk materials (see Fig.~\ref{fig:shear}). The breaking strength of the shear fractures that occurred in the Ta cladding (AS1 and AS2) were slightly above the previously derived USS of Ta (\SI{150}{MPa}). Samples 4S1 (shear fracture at the interface) and 4S2 (Ta foil; shear fracture in the cladding close to the interface) had the same breaking strength (\SI{198}{MPa}) despite presenting different fracture types. Sample 4S1 exhibited a secondary crack inside the Ta2.5W cladding, and both values were close to the previously calculated USS of Ta2.5W (\SI{215}{MPa}). Hence, good diffusion bonding was observed in all TZM shear specimens due to their high interface strengths.

\begin{figure}[htbp]
\centering
\includegraphics[width=\columnwidth]{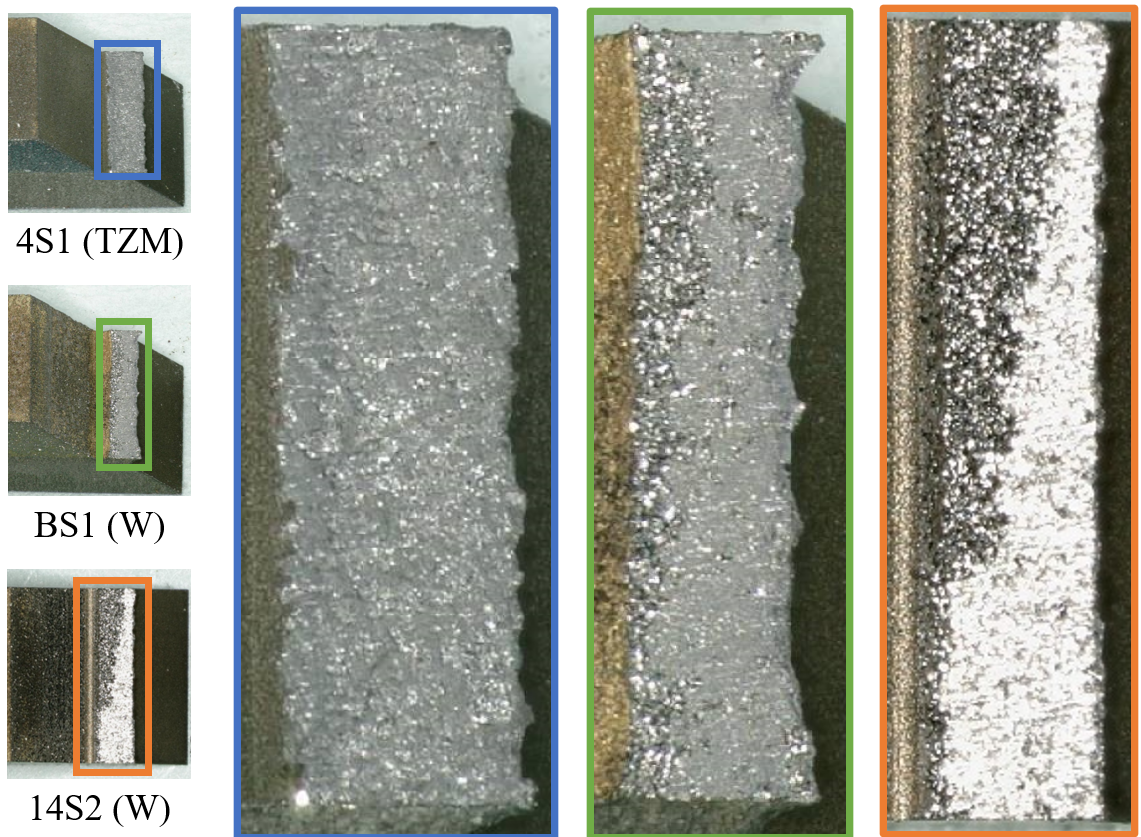}
\caption{Comparison of a shear fracture at the interface of TZM shear specimen 4S1 and the W samples BS1 and 14S2 (Ta foil). The fracture of 4S1 presented a uniform surface, while two separate fracture regions were visible in both BS1 and 14S2.}
\label{fig:fructure_interfaces}
\end{figure}

The conclusions for the W shear specimens were not as clear (see Fig.~\ref{fig:shear}). Half of the specimens (BS2 and 14S1) showed tensile fracture in the W bulk material far away from the interface, indicating interface strengths greater than that of W; however, at 12 and \SI{67}{MPa}, the breaking strengths were unexpectedly low, as the expected UTS of W was \SI{181}{MPa}~\cite{FHinternal2017}. The other two samples (BS1 and 14S2) fractured at the interface, and 14S2 (Ta foil) broke during the final cutting step. The fracture surfaces of BS1 and 14S2 were similar, as two fracture regions were distinguishable; this was not observed in the TZM sample 4S1 (see Fig.~\ref{fig:fructure_interfaces}). It was assumed that patches without diffusion bonding were present in the W samples, causing premature fracture at lower interface strength if the location of these patches correlated with the small interface area of the shear specimens ($1.0 \times 3.6$~mm$^2$). These observations were independent of beam-induced stresses or the presence of Ta foil.

The interface echoes during UT were below the indication threshold for debonding, hence confirming the absence of surface-parallel cracks, delaminations, or volumetric flaws at the interface or in its vicinity. Increased interface echo occurred in the W blocks due to the different reflection coefficients at the interfaces: while 0.4\% of the ultrasonic waves are reflected at a Ta/TZM interface, 7.8\% are reflected at a Ta/W interface, and this can cause higher echo amplitudes.

The thermal-diffusivity specimens 4D (Ta2.5W/TZM) and 14D (Ta/W) showed negligible bonding-interface $R$-values of the order of \SI{1e-6}{m^2\,K\,W^{-1}}. The total thermal resistance of the specimens was mostly dominated by the proportion of the cladding material with respect to the whole sample. The even-numbered samples (with Ta foil) had larger amounts of cladding material due to the presence of the foil, and this increased their total thermal resistance compared to the samples without Ta foil. The results were consistent with those from unirradiated specimens in previous studies~\cite{FHinternal2017, griesemer2024}; therefore, no effects of irradiation or dependency on the proximity to the beam impact were detected from thermal examinations.

The results of EDS mapping for BM1 and the EDS line-scan simulations confirmed successful diffusion bonding. The EDS mapping revealed Ta tips inside the W core, and these matched the shapes of the W grain boundaries. The EDS line scan had a slope that matched the \SI{100}{}-\unit{nm}-diameter beam, although a beam size below \SI{10}{nm} was expected. Hence, mutual diffusion between Ta and W could be derived.

\subsubsection{Ta interlayer as a bonding aid}
The microstructural examination showed no differences in bonding quality between interfaces with and without a Ta interlayer. The limited number of shear specimens was not sufficient to conclude the influence of a Ta interlayer on the bonding strength. The thermal-diffusivity specimens from blocks~4 and 14 did not exhibit differences if the cladding thickness was not increased due to presence of Ta foil. To achieve diffusion bonding between Ta2.5W and W, either an additional Ta interlayer or high HIP parameters (1400\,\unit{\degreeCelsius} and \SI{200}{MPa}) were required. In general, the Ta foil successfully aided the diffusion bonding; however, its use is only recommended when lower HIPing parameters (1200\,\unit{\degreeCelsius} and \SI{150}{MPa}) are applied and not higher ones (1400\,\unit{\degreeCelsius} and \SI{200}{MPa})~\cite{busom2020application}.

\subsection{Pending findings from target-prototype operation}
\subsubsection{Origin of discoloration}
The visually observed deposits on the target were revealed by LOM and EDS analysis. The carbon concentrations measured by EDS in the inner and outer block regions were fairly similar, at approximately \SI{1.13}{wt\,\%} and \SI{1.32}{wt\,\%}, respectively. The deposits consisted largely of C, N, and O and seemed to have weak bonding with the surface as they were removable by C-tape; no other elements were conspicuous in the list of those detected. The presence of carbon in all the irradiated surfaces was likely caused by the degradation of C--N--O-containing substances due to (or during) the beam exposure---possibly residual hydrocarbon molecules---leading to local surface contamination that was correlated with the beam impact~\cite{lopez2024effect,dai2001status}. Therefore, carbon was still regarded as the most likely cause of the discoloration, and further analysis is recommended.

\subsubsection{Heightened temperature in one sensor}
None of the findings from this study can explain the elevated temperature measured by Sensor~8b (see Fig.~\ref{fig:sensor}). In summary, this sensor did not malfunction, the temperature curve measured during beam impact behaved correctly, and the values were physically realistic; however, the cooling rate seemed to be compromised. It is possible that impairment of the cooling rate could have been caused by a thicker layer of cover agent or by compromised water flow due to the presence of sensors on both sides of the water channel. In addition, Sensor~8b was placed slightly closer to the beam-impact location than intended, and the beam was offset toward the sensor. For future instrumentation, it is recommended that temperature sensors should only be installed on one side of a water channel.

\FloatBarrier
\section{Conclusions}
The PIE performed on the prototype BDF target included various destructive and nondestructive testing methods to validate the robustness of the current BDF target design and manufacturing process. These analyses confirmed the absence of any geometrical changes; furthermore, no cracks were present in the cladding material, and no beam-induced effects from temperature gradients or thermally induced stresses were detected. It was determined that there was good bonding quality between the core and cladding materials, and reliable heat transfer from the core to the water-cooling circuit was thus assured. Nevertheless, the sintered W presented high brittleness, having a fully recrystallized microstructure and elevated porosity; this was apparent from superficial cracks detected in the core material. These observations in the W were not considered to be design critical because the appearance of the cracks was considered to be highly consistent with the release of residual stresses during specimen extraction and damage caused to the W microstructure by EDM cutting. To guarantee a longer target lifetime, it is recommended that the manufacturing process should be adapted to obtain a less brittle W material, such as by using hot-rolled W.

The outcome of this analysis will be important for the definition of the final BDF production target, which is scheduled to start operation in 2031. A rebaselining of the project is ongoing, with completion expected in 2026, coinciding with the anticipated approval of the Technical Design Report.

\bibliography{Library}

\end{document}